\documentclass[journal]{IEEEtran}                         

\IEEEoverridecommandlockouts                              

\usepackage{amsfonts}
\usepackage{mathrsfs}
\usepackage{amssymb}
\usepackage{amsmath}
\usepackage{mathrsfs}
\usepackage{array}
\usepackage{extarrows}
\usepackage{color}
\usepackage{cite}
\usepackage{graphicx}
\usepackage[font=footnotesize]{caption}
\usepackage[font=footnotesize]{subcaption}
\usepackage{stfloats}
\usepackage{epstopdf}
\usepackage{arydshln}
\usepackage{ifthen}
\usepackage{url}
\usepackage{pict2e}
\DeclareMathOperator{\grank}{grk \,}
\DeclareMathOperator{\blkdiag}{blkdiag \,}
\DeclareMathOperator{\Spec}{Spec \,}

\DeclareMathOperator{\Image}{Image \,}

\makeatletter 
\newsavebox\myboxA
\newsavebox\myboxB
\newlength\mylenA
\newcommand*\xoverline[2][0.86]{%
    \sbox{\myboxA}{$\m@th#2$}%
    \setbox\myboxB\null
    \ht\myboxB=\ht\myboxA%
    \dp\myboxB=\dp\myboxA%
    \wd\myboxB=#1\wd\myboxA
    \sbox\myboxB{$\m@th\overline{\copy\myboxB}$}
    \setlength\mylenA{\the\wd\myboxA}
    \addtolength\mylenA{-\the\wd\myboxB}%
    \ifdim\wd\myboxB<\wd\myboxA%
       \rlap{\hskip 0.5\mylenA\usebox\myboxB}{\usebox\myboxA}%
    \else
        \hskip -0.5\mylenA\rlap{\usebox\myboxA}{\hskip 0.5\mylenA\usebox\myboxB}%
    \fi}

\makeatletter
\newcommand*\bigcdot{\mathpalette\bigcdot@{.5}}
\newcommand*\bigcdot@[2]{\mathbin{\vcenter{\hbox{\scalebox{#2}{$\m@th#1\bullet$}}}}}
\makeatother

\input{steve.lib}

\title{\LARGE \bf
Structural Completeness of a Multi-channel Linear System with Dependent Parameters
}

\author{F. Liu and A. S. Morse
\thanks{This work was supported by  National Science Foundation grant n. 1607101.00 and  US Air Force grant n. FA9550-16-1-0290.}
\thanks{F. Liu and A. S. Morse are with the Department
of Electrical Engineering, Yale University, New Haven, CT 06511 USA (e-mail: fengjiao.liu@yale.edu; as.morse@yale.edu).}%
\thanks{A preliminary version \cite{liu2019when} of this paper announcing the results but without the proofs of all lemmas was presented at the $58$th IEEE Conference on Decision and Control, Nice, France, December 2019.}%
        }

\begin{document}
\maketitle
\thispagestyle{empty} 
\pagestyle{empty}

\begin{abstract}
It is well known that the ``fixed spectrum" \{i.e., the set of fixed modes\} of a multi-channel linear system plays a central role in the stabilization of such a system with decentralized control. A parameterized multi-channel linear system is said to be ``structurally complete" if it has no fixed spectrum for almost all parameter values. Necessary and sufficient algebraic conditions are presented for a multi-channel linear system with dependent parameters to be structurally complete. An equivalent graphical condition is also given for a certain type of parameterization.
\end{abstract}


\section{Introduction}

The classical ``decentralized control" problem considered in \cite{wang1973stabilization, corfmat1976decentralized} focuses on stabilizing or otherwise controlling a $k>1$ channel linear system of the form
\begin{equation} \label{eqn:syst}
\dot{x} = Ax + \sum_{i=1}^k B_i u_i,\hspace{.3in} y_i = C_i x
\end{equation}
Decentralization is enforced by restricting the feedback of each measured signal $y_i$ to only its corresponding control input $u_i$, possibly through a linear dynamic controller. Wang and Davison \cite{wang1973stabilization} were able to show that no matter what these feedback controllers might be, as long as they are finite dimensional and linear time-invariant (LTI), the spectrum of the resulting closed-loop system contains a fixed subset depending only on $A$, the $B_i$ and the $C_i$, which they elected to called the set of ``fixed modes" of the system. In the sequel, we will use the term ``fixed eigenvalues", because technically modes are not eigenvalues. Roughly speaking, the set of fixed eigenvalues of (\ref{eqn:syst}), henceforth called the ``fixed spectrum" of (\ref{eqn:syst}), is the the spectrum of $A$ that cannot be shifted by the decentralized output feedback laws $u_i = F_i y_i$, $i \in \{1, 2, \dots, k\}$. That is, for given $A \in {\rm I\!R}^{n \times n}$, $B_i \in {\rm I\!R}^{n \times m_i}$, $C_i \in {\rm I\!R}^{l_i \times n}$, the \emph{fixed spectrum} of (\ref{eqn:syst}), written $\Lambda_F$, is precisely
\begin{equation*}
\Lambda_F = \bigcap_{\substack{F_i \in {\rm I\!R}^{m_i \times l_i} \\ i \in \{1, 2, \dots, k\}}} \sigma \left( A+\sum_{i=1}^k B_i F_i C_i \right)
\end{equation*}
where $\sigma(\cdot)$ denotes the spectrum. Since the $F_i$ can be zero, it is clear that the fixed spectrum of (\ref{eqn:syst}) is a subset of the spectrum of $A$. It is possible that the fixed spectrum is an empty set, in which case it is said that the system has no fixed spectrum.

Wang and Davison showed that $\Lambda_F$ is contained in the closed-loop spectrum of the system which results when any given finite dimensional LTI decentralized control is applied to (\ref{eqn:syst}). Thus $\Lambda_F$ must be a stable spectrum if decentralized stabilization is to be achieved with a decentralized LTI control. Wang and Davison also showed that the stability of $\Lambda_F$ is sufficient for decentralized stabilization with linear dynamic controllers. Not surprisingly, the notion of a fixed spectrum arises in connection with the decentralized spectrum assignment problem treated in \cite{corfmat1976decentralized}. In particular it is known that a necessary and sufficient condition for ``free" assignability of an overall closed-loop spectrum with finite dimensional LTI decentralized controllers is that there is no fixed spectrum \cite{wang1973stabilization, corfmat1976decentralized}. However, it should be noted that unlike the centralized case, free spectrum assignability in the decentralized case presumes that the overall spectrum admits a suitable partition into a finite number of symmetric sets, the partition being determined by the strongly connected components in a suitably defined directed graph of (\ref{eqn:syst}) \cite{corfmat1976decentralized}.

It is clear from the preceding that $\Lambda_F$ plays a central role in both the decentralized stabilization and decentralized spectrum assignment problems for a multi-channel linear system. Accordingly many characterizations of $\Lambda_{\text{fixed}}$ exist \cite{gong1992characterization}. Using the canonical form in \cite{morse1973structural}, a test closely related to the existence test for $\Lambda_F$ is given in Corollary 4 of \cite{corfmat1976control}. The idea is further explored and a unifying necessary and sufficient matrix-algebraic condition is proposed for a complex number $\lambda$ to be a fixed eigenvalue \cite{anderson1981algebraic}. Another algebraic condition is established in \cite{gong1994equivalence, liu1996comments}. Equivalent graph-theoretic criteria for (\ref{eqn:syst}) to have a fixed eigenvalue are developed in \cite{reinschke1984graph}. Frequency domain characterizations of $\Lambda_F$ are presented in \cite{seraji1982fixed, anderson1982transfer, vidyasagar1983algebraic, xie1986frequency, xu1986further, kong1996graph}.

While the original work in \cite{wang1973stabilization} defines $\Lambda_F$ and the role it plays in decentralized stabilization, \cite{wang1973stabilization} does not provide a detailed description of $\Lambda_F$ which reflects the fact that different fixed eigenvalues in $\Lambda_F$ may have different properties. In particular, it is known that some of the eigenvalues in $\Lambda_F$ play no significant role in system behavior if one broadens the class of decentralized controllers to include appropriately defined time-varying linear systems \cite{anderson1981time, willems1989time}, or employs other techniques \cite{ozguner1985sampling, willems1988elimination, lavaei2006elimination, stankovic2008stabilization}. The remaining eigenvalues in $\Lambda_F$, called the \emph{firmly fixed eigenvalues}, have been characterized previously in \cite{gong1997stabilization} where they are called the ``quotient fixed modes". The influence of the firmly fixed eigenvalues cannot be avoided with any decentralized control even if it is time-varying and/or nonlinear. The firmly fixed spectrum \{i.e., the set of firmly fixed eigenvalues\} of (\ref{eqn:syst}) is the union of the centralized uncontrollable and/or unobservable spectra of all strongly connected subsystems of (\ref{eqn:syst}) \cite{corfmat1976decentralized, gong1997stabilization}.

As pointed out in \cite{sezer1981structurally}, fixed eigenvalues arise from either one of the following two distinct causes: First, some fixed eigenvalues may be present due to exact matchings of certain nonzero entries in various locations in system coefficient matrices $A$, the $B_i$ and the $C_i$, so slight independent perturbations of these nonzero entries might circumvent such fixed eigenvalues. Second, other fixed eigenvalues may be a consequence of certain zero and nonzero patterns of entries in the system coefficient matrices, thus the existence of such fixed eigenvalues is intrinsic in the sense that they cannot be avoided by zero/nonzero pattern-preserving perturbations, though their values may vary with these perturbations. Just to clarify, a fixed eigenvalue of the second type and a firmly fixed eigenvalue are two different concepts; a fixed eigenvalue of the second type may or may not be a firmly fixed eigenvalue. In models of real physical systems, parameter values are usually not known exactly, thus in such cases it is unlikely that there will be perfect matchings of system parameter values. For this reason, the existence of fixed eigenvalues of the second type is the main concern of practical importance. Fixed eigenvalues arising in this manner were first studied by Sezer and {\v{S}}iljak in 1981 \cite{sezer1981structurally} and were called ``structurally fixed modes" \cite{siljak1991decentralized}. The term ``structural" originates from the concept of structural controllability introduced by Lin in 1974 \cite{lin1974structural}, who assumed that each entry of the system coefficient matrices is either a fixed zero or a distinct scalar parameter and all parameters are algebraically independent. However, unlike fixed eigenvalues, ``structurally fixed modes" and the number of them are in general not numerically ``fixed" but functions of the parameters or the zero pattern-preserving perturbations. To avoid confusion, it makes more sense to define a new property of the system rather than to adopt the term ``structurally fixed eigenvalues" or ``structurally fixed spectrum". This new property is called ``structural completeness", where the term ``completeness" comes from \cite{corfmat1976control, corfmat1976decentralized} and is generalized a little bit (by allowing a zero transfer matrix) in the context of this paper. Accordingly, a linear system of the form (\ref{eqn:syst}) whose coefficient matrices $A$, the $B_i$ and the $C_i$ depend algebraically on a vector $p$ of parameters, is said to be \emph{structurally complete} if it has no fixed spectrum for some value of $p$. The algebraic condition in \cite{anderson1981algebraic} (restated as Proposition \ref{prp:mtrx-pencil} in Section \ref{sec:analyses} of this paper) and Lemma \ref{lem:partition} in this paper clearly suggest that the set of values of $p$ for which such a parameterized linear system has no fixed spectrum is either an empty set or the complement of a proper algebraic set in the parameter space. Thus if such a system is structurally complete, it has no fixed spectrum for almost every value of $p$; if not, it has a fixed spectrum for each fixed value of $p$ and of course the fixed spectrum may depend on $p$.

With Lin's assumption of algebraically independent nonzero entries, Sezer and {\v{S}}iljak derived necessary and sufficient matrix-algebraic conditions for a linear system of the form (\ref{eqn:syst}) to be structurally complete. An equivalent but less explicit algebraic condition was provided in 1983 \cite{momen1983structurally}. The algebraic conditions in \cite{sezer1981structurally} were soon converted to equivalent graph-theoretic conditions in \cite{linnemann1983fixed, pichai1984graph, papadimitriou1984simple}. Based on the graphical conditions, some design problems with the requirement of structural completeness are considered in \cite{trave1987minimal, pequito2013designing}.

However, Lin's assumption that each nonzero entry in the system coefficient matrices is a distinct scalar parameter is not applicable to systems in which a parameter may appear in multiple locations of the system coefficient matrices. Therefore, there is need to study the genericity of a fixed spectrum using more general types of parameterizations, such as ``linear parameterizations" \cite{corfmat1976structurally, liu2019graphical}, or more general parameterizations in which the nonzero entries of the system coefficient matrices are polynomials in the parameters of interest. This is what this paper does.

The rest of the paper is organized as follows. Three more general types of parameterizations are defined and the problems studied in this paper are formulated in Section \ref{sec:prob-formulation}. Some terminology and concepts are defined in Section \ref{sec:preliminaries}. The main results of this paper are presented in Section \ref{sec:results} and proved in Section \ref{sec:analyses}.

\section{Problem Formulation} \label{sec:prob-formulation}

First, various parameterizations of multi-channel linear systems are defined in the order from the most general to the most specific. Then the problems regarding the parameterizations are formulated in the last paragraph of this section.

Let $p \in {\rm I\!R}^q$ be a vector of $q > 0$ algebraically independent parameters $p_1$, $p_2$, $\dots$, $p_q$. A $k$-channel linear system $\{A(p), B_i(p), C_i(p); k\}$ is \emph{polynomially parameterized} if all the entries of its coefficient matrices are in the polynomial ring ${\rm I\!R}[p_1, p_2, \dots, p_q]$. A good example of a polynomially parameterized multi-channel system is given by Example 1.36 on page 33 of \cite{siljak1991decentralized}, in which a system of two identical inverted pendulums coupled by a spring is modeled with 
\begin{align} \label{exp:poly-pendulum}
&\dot{x} = 
\begin{bmatrix}
0 & 1 & 0 & 0 \\
\frac{g}{l} - \frac{ka^2}{ml^2} & 0 & \frac{ka^2}{ml^2} & 0 \\
0 & 0 & 0 & 1 \\
\frac{ka^2}{ml^2} & 0 & \frac{g}{l} - \frac{ka^2}{ml^2} & 0 
\end{bmatrix}
\! x
+ \!
\begin{bmatrix}
0 \\
\frac{1}{ml^2} \\
0 \\
0
\end{bmatrix}
\! u_1
+ \!
\begin{bmatrix}
0 \\
0 \\
0 \\
\frac{1}{ml^2}
\end{bmatrix}
\! u_2 \nonumber \\
&y_1 = 
\begin{bmatrix}
1 & 0 & 0 & 0
\end{bmatrix}
x
, \hspace{7mm}
y_2 = 
\begin{bmatrix}
0 & 0 & 1 & 0
\end{bmatrix}
x
\end{align}
where $g$ is the local acceleration of gravity, $m$ and $l$ are respectively the mass and the lengths of the two pendulums, $k$ is the stiffness of the spring, and $a$ is the length measured from the pivot point of a pendulum to the point on the pendulum rod to which the spring is attached. Due to inevitable measurement errors, the exact values of these physical characteristics of the system may be unknown. Let $p_1 = g$, $p_2 = \frac{1}{m}$, $p_3 = \frac{1}{l}$, and $p_4 = ka^2$, then the 2-channel system (\ref{exp:poly-pendulum}) is polynomially parameterized in the four algebraically independent parameters $p_1$ through $p_4$.

Linear parameterization is a special case of polynomial parameterization, which addresses some simple but commonly encountered modeling situations such as when $A$, $b_1$, $b_2$, $c_1$ and $c_2$ are of the forms
\begin{align} \label{eqn:lnr-param-expl}
& A = 
\begin{bmatrix}
p_1 & p_1 \\
0   & p_2
\end{bmatrix}
, \quad
b_1 = 
\begin{bmatrix}
0 \\
p_2
\end{bmatrix}
, \quad
b_2 = 
\begin{bmatrix}
p_3 \\
0
\end{bmatrix}
\nonumber \\
& c_1 = 
\begin{bmatrix}
p_4 & 0
\end{bmatrix}
, \quad
c_2 = 
\begin{bmatrix}
p_1 & p_1
\end{bmatrix}
\end{align}
where at least one parameter, in this example $p_1$ and $p_2$, appears in more than one location. Let
\begin{equation} \label{eqn:stackmatrix}
B_{n \times m} \triangleq [B_1 \enspace B_2 \enspace \dots \enspace B_k], \qquad 
C_{l \times n} \triangleq 
\begin{bmatrix}
C_1 \\
C_2 \\
\vdots \\
C_k
\end{bmatrix}
\end{equation}
where $m = \sum_{i=1}^k m_i$ and $l = \sum_{i=1}^k l_i$. A $k$-channel linear system $\{A(p), B_i(p), C_i(p); k\}$ is \emph{linearly parameterized} if the partial derivative of the block partitioned matrix 
$\begin{bmatrix}
A & B \\
C & \mathbf{0}
\end{bmatrix}$
with respect to each parameter is a rank-one matrix, where $B$ and $C$ are given by (\ref{eqn:stackmatrix}). That is, a linearly parameterized $k$-channel system $\{A(p), B_i(p), C_i(p); k\}$ can be written as
\begin{equation} \label{eqn:lnr-param-def}
\begin{bmatrix}
A_{n \times n}(p)  &  B_{n \times m}(p)  \\
C_{l \times n}(p)  &  \mathbf{0}
\end{bmatrix}
=
\sum_{i \in \mathbf{q}} g_i p_i h_i
\end{equation}
where $\mathbf{q} \triangleq \{1, 2, \dots, q\}$, and for each $i \in \mathbf{q}$, $g_i \in {\rm I\!R}^{n+l}$, $h_i \in {\rm I\!R}^{1 \times (n+m)}$. Note that equation (\ref{eqn:lnr-param-def}) implies that a parameter cannot appear in both $B$ and $C$, otherwise the lower right block of the partitioned matrix on the left-hand side of (\ref{eqn:lnr-param-def}) would be nonzero. As in \cite{liu2019graphical}, it will be assumed for simplicity and without loss of generality that the set of matrices $\{g_1 h_1, g_2 h_2, \dots, g_q h_q\}$ is linearly independent. This implies that $q \leq n(n+m+l)$.

Before proceeding, we point out that not every system $\{A(p), B_i(p), C_i(p); k\}$ with parameters entering its coefficient matrices ``linearly" is a linear parameterization as defined here. For example, while the system shown in (\ref{eqn:lnr-param-expl}) is linearly parameterized, the system
\begin{align} \label{eqn:lnr-param-cntexpl}
& A = 
\begin{bmatrix}
p_1 & 0 \\
0   & p_1
\end{bmatrix}
, \quad
b_1 = 
\begin{bmatrix}
0 \\
p_2
\end{bmatrix}
, \quad
b_2 = 
\begin{bmatrix}
p_3 \\
0
\end{bmatrix}
\nonumber \\
& c_1 = 
\begin{bmatrix}
p_4 & 0
\end{bmatrix}
, \quad
c_2 = 
\begin{bmatrix}
p_1 & p_1
\end{bmatrix}
\end{align}
is not. The same argument in \cite{liu2019graphical} applies here that a $k$-channel linear system $\{A(p), B_i(p), C_i(p); k\}$ of which the entries of the coefficient matrices depend linearly on $q$ parameters $p_1$, $p_2$, $\dots$, $p_q$ will be linearly parameterized if and only if all minors of the partitioned matrix 
$\begin{bmatrix}
A \quad B \\
C \quad \mathbf{0}
\end{bmatrix}$ are multilinear functions of the $q$ parameters. It is clear that the matrices in (\ref{eqn:lnr-param-cntexpl}) do not have this property.

Some common scenarios of linear parameterization arise in multi-agent networks, flow networks, resistor networks, and spring networks, whose dynamics are characterized by the Laplacian matrices\footnote{A standard Laplacian matrix is defined as $L = D - J$, where $D$ is the degree matrix and $J$ is the adjacency matrix of the network, while a signless Laplacian matrix is defined as $L = D + J$.} of the networks \cite{saber2003consensus, zamani2009structural, spielman2015spectral}. Suppose on such a network some vertices are controlled by external inputs, the states of some vertices are measured as output signals, and the output feedback structure satisfies a certain decentralized constraint, then the network can be modeled with a decentralized multi-channel system. If each nonzero off-diagonal entry of the (possibly signless) Laplacian matrix of this network is represented by a distinct parameter, the system is linearly parameterized, for the diagonal entries of the Laplacian matrix are linear combinations of the off-diagonal entries in the same row.

The linear parameterization defined above is said to satisfy the \emph{binary assumption} if all of the $g_i$ and $h_i$ appearing in (\ref{eqn:lnr-param-def}) are binary vectors, i.e., vectors of 1's and 0's. So the binary assumption requires that all nonzero coefficients of the parameters are 1's, which is a special case of linear parameterization. As a quick example, if all edges in a multi-agent network described in the paragraph above have negative signs, the resulting Laplacian matrix is signless, thus the network can be written as a linearly parameterized multi-channel system which satisfies the binary assumption. Similarly, a linear parameterization satisfies the \emph{unitary assumption} if all of the $g_i$ and $h_i$ appearing in (\ref{eqn:lnr-param-def}) are unit vector, i.e., vectors with $1$ in one entry and $0$ in all other entries. The unitary assumption is clearly a special case of the binary assumption. Note that Lin's assumption is exactly the linear parameterization satisfying the unitary assumption. The relations between all the parameterizations studied in this paper are summarized below. 
\begin{align*}
&\text{Lin's assumption} \\
= \enspace & \text{linear parameterization satisfying the unitary assumption} \\
\subset \enspace & \text{linear parameterization satisfying the binary assumption} \\
\subset \enspace & \text{linear parameterization} \\
\subset \enspace & \text{polynomial parameterization} 
\end{align*}

Let $F_{m \times l} \triangleq \blkdiag \{F_1, F_2, \dots, F_k\}$ be a block diagonal matrix with $\tilde{q} \triangleq \sum_{i=1}^k m_i l_i$ nonzero entries. Each of these nonzero entries can be represented by a distinct parameter $\tilde{p}_i$, then the resulting parameterized block diagonal matrix is denoted by $F(\tilde{p})$, where $\tilde{p} \in {\rm I\!R}^{\tilde{q}}$ is a vector of $\tilde{q}$ algebraically independent parameters $\tilde{p}_1$, $\tilde{p}_2$, $\dots$, $\tilde{p}_{\tilde{q}}$. Note that $F(\tilde{p})$ is linearly parameterized and satisfies the unitary assumption.

With the parameterizations defined above, the problem of interest is to find conditions for the existence of a parameter vector $p \in {\rm I\!R}^q$ for which a parameterized $k$-channel system $\{A(p), B_i(p), C_i(p); k\}$ has no fixed spectrum. If such values exist, the parameterized system is \emph{structurally complete}. Such a polynomially parameterized $k$-channel system has no fixed spectrum for almost every value of $p \in {\rm I\!R}^q$ in the sense that the set of values of $p \in {\rm I\!R}^q$ for which the system has no fixed spectrum is the complement of a proper algebraic set in ${\rm I\!R}^q$. The system is said to be \emph{structurally incomplete} if it is not structurally complete. Thus if a polynomially parameterized system $\{A(p), B_i(p), C_i(p); k\}$ is structurally incomplete, it has a fixed spectrum for each fixed $p \in {\rm I\!R}^q$ and the fixed spectrum is a function of $p$. Note that there always exists $p \in {\rm I\!R}^q$ such that a polynomially parameterized system has a fixed spectrum, for which a pathological case is that when $p = 0$, the system has a fixed spectrum consisting solely of 0's.

This paper gives necessary and sufficient matrix-algebraic conditions for a polynomially parameterized $k$-channel system and a linearly parameterized $k$-channel system, respectively, to be structurally complete. This paper also provides an equivalent graph-theoretic condition for a linearly parameterized system which satisfies the binary assumption. To the best of our knowledge, these algebraic and graphical conditions are the first results on more general types of parameterizations that allow a parameter to appear in multiple system coefficient matrices.

\section{Preliminaries} \label{sec:preliminaries}

In order to state the main results of this paper, some terminology and a number of graphical concepts are needed.

The \emph{generic rank} of a polynomially parameterized matrix $M$, denoted by $\grank M$, is the maximum rank of $M$ that can be achieved as the parameters vary over the entire parameter space. It is generic in the sense that it is achievable by any parameter values in the complement of a proper algebraic set in the parameter space. For example, $\grank (A(p)+B(p)F(\tilde{p})C(p))$ is the the maximum rank of $A+BFC$ that can be achieved as $p$ varies over ${\rm I\!R}^q$ and $\tilde{p}$ varies over ${\rm I\!R}^{\tilde{q}}$, and it is achievable by almost any $p$ and $\tilde{p}$ in ${\rm I\!R}^q \times {\rm I\!R}^{\tilde{q}}$, where $\times$ denotes the Cartesian product. Note that the generic rank of $A(p)+B(p)F(\tilde{p})C(p)$ depends only on the parameterized system $\{A(p), B_i(p), C_i(p); k\}$, since $F(\tilde{p})$ is determined solely by the dimensions of the $B_i(p)$ and the $C_i(p)$.

Let $(C, A, B)$ be a real matrix triple. Let $\mathscr{B}$ denote the image of $B$. Let 
\begin{equation*}
\left\langle A \, | \, \mathscr{B} \right\rangle \triangleq \mathscr{B} + A\mathscr{B} + A^2\mathscr{B} + \dots + A^{n-1}\mathscr{B}
\end{equation*} 
be the \emph{controllable space} of $(A, B)$ and let
\begin{equation*}
[C \, | \, A] \triangleq \bigcap_{i=1}^{n} \ker (CA^{i-1})
\end{equation*} 
be the \emph{unobservable space} of $(C, A)$. Let $\mathbf{k} \triangleq \{1, 2, \dots, k\}$. Suppose $\mathcal{S} = \{i_1, i_2, \dots, i_s\} \subset \mathbf{k}$ with $i_1 < i_2 < \dots < i_s$, the complement of $\mathcal{S}$ in $\mathbf{k}$ is denoted by $\mathbf{k} - \mathcal{S} = \{j_1, j_2, \dots, j_{k-s}\}$ with $j_1 < j_2 < \dots < j_{k-s}$. Let 
\begin{equation*}
B_{\mathcal{S}} \triangleq [B_{i_1} \enspace B_{i_2} \enspace \dots \enspace B_{i_s}], \qquad
C_{\mathbf{k} - \mathcal{S}} \triangleq 
\begin{bmatrix}
C_{j_1} \\
C_{j_2} \\
\vdots \\
C_{j_{k-s}}
\end{bmatrix}
\end{equation*}
Similarly, let $\mathscr{B}_{\mathcal{S}}$ denote the image of $B_{\mathcal{S}}$, let $\left\langle A \, | \, \mathscr{B}_{\mathcal{S}} \right\rangle$ denote the controllable space of $(A, B_{\mathcal{S}})$, and let $[C_{\mathbf{k} - \mathcal{S}} \, | \, A]$ denote the unobservable space of $(C_{\mathbf{k} - \mathcal{S}}, A)$. By convention, $\left\langle A \, | \, \mathscr{B}_{\emptyset} \right\rangle = \emptyset$ and $[C_{\emptyset} \, | \, A] = {\rm I\!R}^n$. Given two subspaces $\mathscr{X}_1, \mathscr{X}_2 \subset {\rm I\!R}^n$, if $\mathscr{X}_1 \subset \mathscr{X}_2$ and $\mathscr{X}_1 \neq \mathscr{X}_2$, $\mathscr{X}_1$ is called a \emph{proper subspace} of $\mathscr{X}_2$.

A \emph{strongly connected component} of a directed graph is a maximal subgraph subject to being strongly connected\footnote{A directed graph is \emph{strongly connected} if there is a path from every vertex to every other vertex.}. The collection of strongly connected components of a directed graph forms a partition of its vertex set. A \emph{directed cycle graph} is a strongly connected graph whose vertices can be labeled in the order $1$ to $t$ for some $t \in {\rm I\!N}$ such that the arcs are $(i, i+1)$ and $(t, 1)$, where $i=1, 2, \dots, t-1$. So in a directed cycle graph, each vertex has exactly one incoming arc and one outgoing arc. One vertex with a single self-loop is also a directed cycle graph. As this paper is concerned with directed graphs only, a directed cycle graph will be simply called a cycle graph in the rest of the paper. The \emph{disjoint union} of two or more graphs is the union of these graphs whose vertex sets are disjoint.

The \emph{graph} of a linearly parameterized $k$-channel system $\{A(p), B_i(p), C_i(p); k\}$, written $\mathbb{G} = \{\mathcal{V}, \mathcal{E}\}$, is defined to be an unweighted directed multigraph\footnote{A multigraph is a graph that allows parallel arcs and self-loops.} with vertex set $\mathcal{V}$ and arc set $\mathcal{E}$. With a slight abuse of notation, let $x_i$, $u_i$, and $y_i$ denote a state vertex, an input vertex, and an output vertex, respectively. Let $\mathcal{V}_x \triangleq \{x_1, x_2, \dots, x_n\}$ be the set of \emph{state vertices}, one vertex for each state variable. Let $\mathcal{V}_u \triangleq \{u_1, u_2, \dots, u_m\}$ be the set of \emph{input vertices}, one vertex for each input. Let $\mathcal{V}_y \triangleq \{y_1, y_2, \dots, y_l\}$ be the set of \emph{output vertices}, one vertex for each output. Then the vertex set 
\begin{equation*}
\mathcal{V} \triangleq \mathcal{V}_x \cup \mathcal{V}_u \cup \mathcal{V}_y
\end{equation*}
which has $n+m+l$ vertices. Each arc in $\mathbb{G}$ has a color associated with it, indicating the parameter that attributes to this arc. In the sequel, $(v_i, v_j)_r$ denotes an arc from $v_i$ to $v_j$ with color\footnote{In this paper, each color is labeled by a distinct integer.} $r$, where $v_i, v_j \in \mathcal{V}$. Let 
\begin{align*}
\mathcal{E}_A &\triangleq \{(x_j, x_i)_r \, | \, \text{the} \enspace ij\text{th entry of} \enspace A(p) \enspace \text{contains} \enspace p_r; \\
& \hspace{2.32in} x_j, x_i \in \mathcal{V}_x\} \\
\mathcal{E}_B &\triangleq \{(u_j, x_i)_r \, | \, \text{the} \enspace ij\text{th entry of} \enspace B(p) \enspace \text{contains} \enspace p_r; \\
& \hspace{2in} u_j \in \mathcal{V}_u, x_i \in \mathcal{V}_x\} \\
\mathcal{E}_C &\triangleq \{(x_j, y_i)_r \, | \, \text{the} \enspace ij\text{th entry of} \enspace C(p) \enspace \text{contains} \enspace p_r; \\
& \hspace{2.02in} x_j \in \mathcal{V}_x, y_i \in \mathcal{V}_y\}
\end{align*}
Then the arc set 
\begin{equation*}
\mathcal{E} \triangleq \mathcal{E}_A \cup \mathcal{E}_B \cup \mathcal{E}_C
\end{equation*}
Graph $\mathbb{G}$ has $q$ colors, as there are $q$ parameters in $A(p)$, $B(p)$, and $C(p)$.

However, graph $\mathbb{G}$ does not tell which input or output vertex belongs to which channel, thus does not show which is the allowed configuration for decentralized control. As the pattern of nonzero entries in the block diagonal matrix $F(\tilde{p})$ reveals the allowed configuration for decentralized output feedback, it is desirable to have a graph capturing this allowed configuration. Because each nonzero entry of $F(\tilde{p})$ is a distinct parameter $\tilde{p}_i$ and corresponds to an arc from an output vertex to an input vertex, there are $\tilde{q}$ allowed arcs for decentralized output feedback, each of which has a distinct color from the $q$ colors in $\mathbb{G}$. Let 
\begin{multline*}
\mathcal{E}_F \triangleq \{(y_j, u_i)_{\tilde{r}} \, | \, \text{the} \enspace ij\text{th entry of} \enspace F(\tilde{p}) \enspace \text{contains} \enspace \tilde{p}_{\tilde{r}}; \\ 
y_j \in \mathcal{V}_y, u_i \in \mathcal{V}_u \} 
\end{multline*}
It is clear that $|\mathcal{E}_F| = \tilde{q}$, where $|\bigcdot|$ denotes the cardinality of a set. The \emph{feedback graph} $\mathbb{G}_F$ of a linearly parameterized $k$-channel system $\{A(p), B_i(p), C_i(p); k\}$ with decentralized output feedback matrices $\{F_i(\tilde{p}); k\}$ is defined as 
\begin{equation*}
\mathbb{G}_F \triangleq \{\mathcal{V}, \mathcal{E} \cup \mathcal{E}_F\}
\end{equation*}
Graph $\mathbb{G}_F$ has $q + \tilde{q}$ colors, as there are $q + \tilde{q}$ parameters in $A(p)$, $B(p)$, $C(p)$, and $F(\tilde{p})$. It is worth pointing out that although the feedback graph $\mathbb{G}_F$ seemingly depends on both the $k$-channel system and the block diagonal matrix $F(\tilde{p})$, in fact $\mathbb{G}_F$ is uniquely determined only by the $k$-channel system, for matrix $F(\tilde{p})$ is determined solely by the dimensions of the $B_i(p)$ and the $C_i(p)$.

\begin{figure}[!h]
    \centering
    \includegraphics[width=0.3\textwidth]{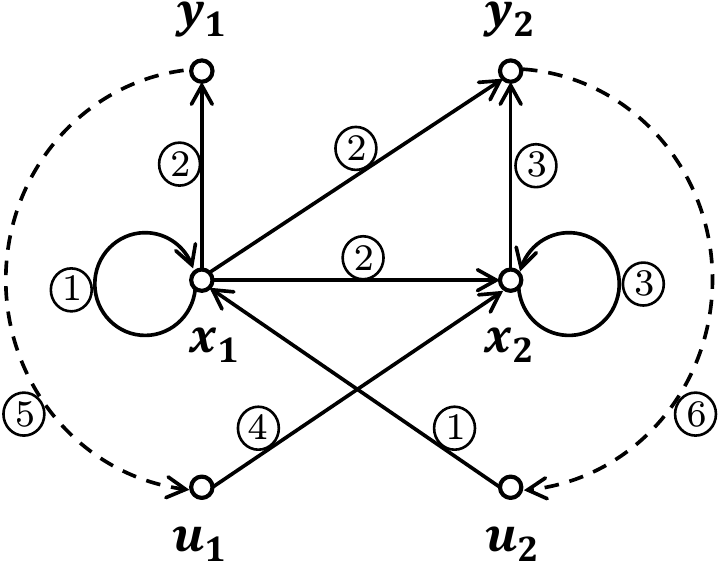}
    \caption{The feedback graph $\mathbb{G}_F$ of the 2-channel system in (\ref{eqn:graph-expl}).}
    \label{fig:Graph-F}
\end{figure}

Figure \ref{fig:Graph-F} shows the feedback graph $\mathbb{G}_F$ of a 2-channel system
\begin{align} \label{eqn:graph-expl}
& A = 
\begin{bmatrix}
p_1 & 0 \\
p_2 & p_3
\end{bmatrix}
, \quad
b_1 = 
\begin{bmatrix}
0 \\
p_4
\end{bmatrix}
, \quad
b_2 = 
\begin{bmatrix}
p_1 \\
0
\end{bmatrix}
\nonumber \\
& c_1 = 
\begin{bmatrix}
p_2 & 0
\end{bmatrix}
, \quad
c_2 = 
\begin{bmatrix}
p_2 & p_3
\end{bmatrix}
\end{align}
where the arcs in $\mathcal{E}$ are drawn in solid lines, the arcs in $\mathcal{E}_F$ are drawn in dashed lines, and symbol \textcircled{\footnotesize{$k$}} labels color $k$ for $k = 1, 2, \dots, 6$.

\begin{figure*}[!b]
	\centering
	\begin{subfigure}{0.3\textwidth}
		\includegraphics[width=\linewidth]{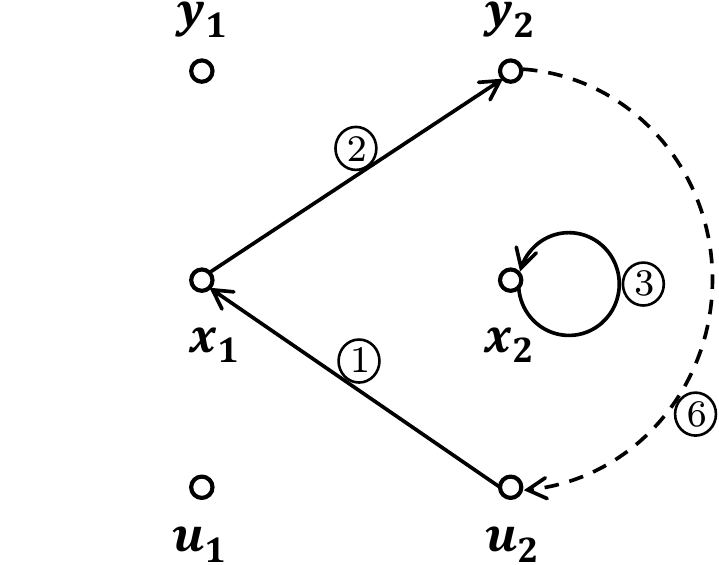}
		\caption{An even multi-colored cycle subgraph.} 
		\label{subfig:even-cycle-similar}
	\end{subfigure}
	\hfil  \hfil  \hfil  \hfil  
	\begin{subfigure}{0.3\textwidth}
		\includegraphics[width=\linewidth]{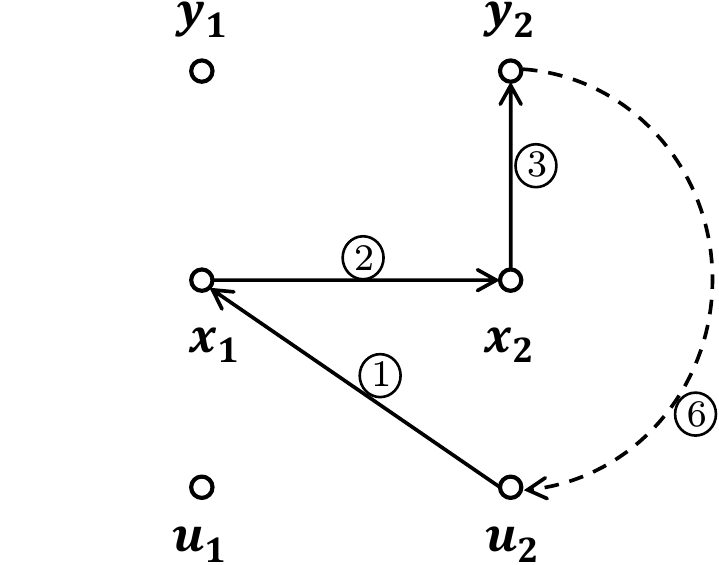}
		\caption{An odd multi-colored cycle subgraph.} 
		\label{subfig:odd-cycle-similar}
	\end{subfigure}
	\caption{A balanced similarity class of multi-colored cycle subgraphs of the feedback graph in Figure \ref{fig:Graph-F}.} 	
	\label{fig:even-odd-cycle-similar}
\end{figure*}

\begin{figure*}[!b]
	\centering
	\begin{subfigure}{0.3\textwidth}
		\includegraphics[width=\linewidth]{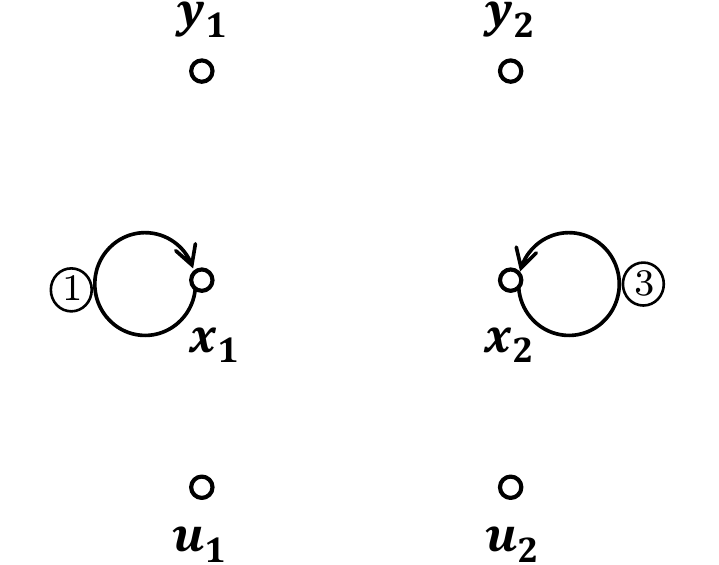}
		\caption{An even multi-colored cycle subgraph.} 
		\label{subfig:even-cycle}
	\end{subfigure}
	\hfil  \hfil  \hfil  \hfil  
	\begin{subfigure}{0.3\textwidth}
		\includegraphics[width=\linewidth]{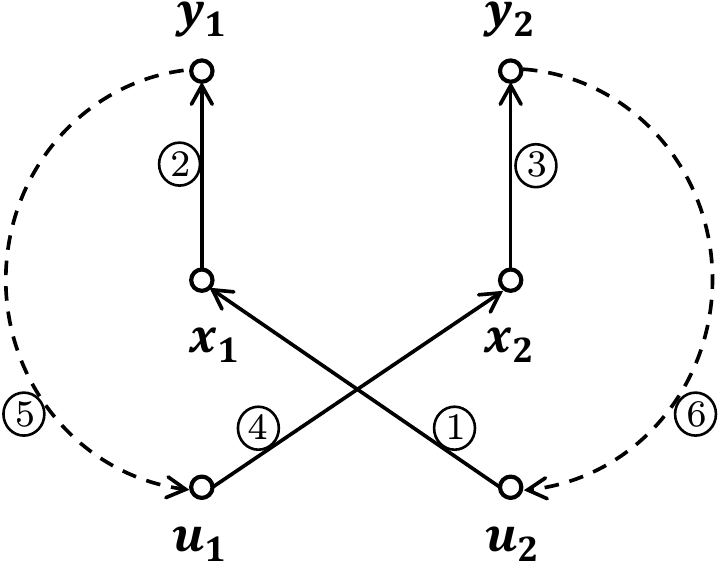}
		\caption{An odd multi-colored cycle subgraph.} 
		\label{subfig:odd-cycle}
	\end{subfigure}
	\caption{Two unbalanced similarity classes of multi-colored cycle subgraphs of the feedback graph in Figure \ref{fig:Graph-F}.} 	
	\label{fig:even-odd-cycle}
\end{figure*}

Note that a feedback graph $\mathbb{G}_F$ has four properties: (\lowercase\expandafter{\romannumeral1}) Input vertices have incoming arcs only from output vertices and have outgoing arcs only to state vertices. Similarly, output vertices have incoming arcs only from state vertices and have outgoing arcs only to input vertices. (\lowercase\expandafter{\romannumeral2}) An arc in $\mathcal{E}_A$ and an arc in $\mathcal{E}_B$ may have the same color. An arc in $\mathcal{E}_A$ and an arc in $\mathcal{E}_C$ may have the same color. But an arc in $\mathcal{E}_B$ and an arc in $\mathcal{E}_C$ never share the same color, as a parameter never appears in both $B(p)$ and $C(p)$. Each arc in $\mathcal{E}_F$ has a distinct color from all the colors in $\mathcal{E}$. (\lowercase\expandafter{\romannumeral3}) In $\mathcal{E}$, there may be more than one arc from one given vertex $j$ to another vertex $i$, for the corresponding entry of the system coefficient matrices may be a linear combination of more than one parameter. If this is the case, all arcs from vertex $j$ to vertex $i$ will have distinct colors. In $\mathcal{E}_F$, there are no parallel arcs. (\lowercase\expandafter{\romannumeral4}) In $\mathcal{E}_A \cup \mathcal{E}_B$ (respectively, $\mathcal{E}_A \cup \mathcal{E}_C$), if there are two arcs of color $r \in \mathbf{q}$, one leaving vertex $j$ and the other pointing toward vertex $i$, then there must be an arc $(v_j, v_i)_r$ in $\mathcal{E}_A \cup \mathcal{E}_B$ (respectively, $\mathcal{E}_A \cup \mathcal{E}_C$). This is due to the rank-one constraint for each parameter in linear parameterization.

As noted before, a feedback graph $\mathbb{G}_F$ is uniquely determined by a linearly parameterized $k$-channel system. However, from $\mathbb{G}_F$ one cannot recover the linearly parameterized system. This is because $\mathbb{G}_F$ is unweighted, which means it cannot reflect the coefficients or the specific functions of the parameters appearing in the nonzero entries of the matrices\footnote{By introducing ``weights" on $\mathbb{G}_F$, one can uniquely identify a linearly parameterized $k$-channel system from a weighted $\mathbb{G}_F$.}. Nevertheless, from a feedback graph $\mathbb{G}_F$ one can write down a unique linearly parameterized $k$-channel system which satisfies the binary assumption. So it is possible to characterize such a system solely in terms of its feedback graph. Therefore, this paper deals exclusively with graphical characterizations of linearly parameterized systems which satisfy the binary assumption.

A \emph{multi-colored cycle subgraph} of a feedback graph $\mathbb{G}_F$ is a subgraph of $\mathbb{G}_F$, which is the disjoint union of a finite number of cycle graphs with all state vertices contained in the union graph and with each arc in the union graph of a different color. Let $\mathcal{C}(\mathbb{G}_F)$ denote the set of all multi-colored cycle subgraphs of $\mathbb{G}_F$. Two multi-colored cycle subgraphs $\mathbb{S}_1, \mathbb{S}_2 \in \mathcal{C}(\mathbb{G}_F)$ are called \emph{similar} if $\mathbb{S}_1$ and $\mathbb{S}_2$ have the same set of colors. Note that similar multi-colored cycle subgraphs also have the same number of arcs. Graph similarity is an equivalence relation on $\mathcal{C}(\mathbb{G}_F)$. The corresponding equivalence classes induced by this relation are called \emph{similarity classes}. A multi-colored cycle subgraph is \emph{odd} (respectively, \emph{even}) if it has an odd (respectively, even) number of cycle graphs. A similarity class of multi-colored cycle subgraphs is \emph{balanced} if the numbers of odd and even multi-colored cycle subgraphs in the similarity class are equal. Otherwise, it is \emph{unbalanced}.

To illustrate these concepts, let $\mathbb{G}_F$ be the feedback graph in Figure \ref{fig:Graph-F}. Then $\mathbb{G}_F$ has four multi-colored cycle subgraphs, as shown in Figure \ref{fig:even-odd-cycle-similar} and Figure \ref{fig:even-odd-cycle}. The two graphs in Figure \ref{fig:even-odd-cycle-similar} are in the same similarity class with colors 1, 2, 3, 6. As the graph in Figure \ref{subfig:even-cycle-similar} is even and the graph in Figure \ref{subfig:odd-cycle-similar} is odd, this similarity class is balanced. On the other hand, each of the two graphs in Figure \ref{fig:even-odd-cycle} forms its own similarity class, which is unbalanced. Thus the feedback graph in Figure \ref{fig:Graph-F} has a balanced similarity class and two unbalanced ones.

\section{Main Results} \label{sec:results}

The following three theorems give necessary and sufficient conditions for the structural completeness of three types of parameterized multi-channel linear systems, respectively.

\begin{theorem} \label{thm:poly}
A polynomially parameterized $k$-channel system $\{A(p), B_i(p), C_i(p); k\}$ is structurally complete if and only if $\forall \mathcal{S} \subset \mathbf{k}$, $\exists p \in {\rm I\!R}^q$ such that 
\begin{equation} \label{eqn:poly-partition}
\rank 
\begin{bmatrix}
\lambda I_n -A & B_{\mathcal{S}} \\
C_{\mathbf{k}-\mathcal{S}} & \mathbf{0}
\end{bmatrix}
\geq n \quad \text{for every} \enspace \lambda \in \sigma(A)
\end{equation}
or equivalently, 
\begin{equation} \label{eqn:poly-fixed-eigval}
\bigcap_{\substack{E_i \in {\rm I\!R}^{m_i \times n}, \, i \in \mathcal{S} \\ K_j \in {\rm I\!R}^{n \times l_j}, \, j \in \mathbf{k}-\mathcal{S}}} \sigma \left( A + \sum_{i \in \mathcal{S}} B_i E_i + \sum_{j \in \mathbf{k}-\mathcal{S}} K_j C_j \right) = \emptyset
\end{equation}
\end{theorem}

\begin{remark} \label{rmk:poly}
If the condition in Theorem \ref{thm:poly} is not satisfied, i.e., $\exists \mathcal{S} \subset \mathbf{k}$ such that $\forall p \in {\rm I\!R}^q$, (\ref{eqn:poly-partition}) does not hold, and furthermore, if the transfer matrix $C_{\mathbf{k} - \mathcal{S}}(p) (\lambda I - A(p))^{-1} B_{\mathcal{S}}(p) \equiv 0$, then for each fixed $p \in {\rm I\!R}^q$, $\left\langle A \, | \, \mathscr{B}_{\mathcal{S}} \right\rangle$ is a proper subspace of $[C_{\mathbf{k} - \mathcal{S}} \, | \, A]$ and any eigenvalue of the map induced in the quotient space \cite{wonham1985linear} $[C_{\mathbf{k} - \mathcal{S}} \, | \, A] / \left\langle A \, | \, \mathscr{B}_{\mathcal{S}} \right\rangle$ by $A$ is a firmly fixed eigenvalue of the system. Of course the numerical values of the firmly fixed eigenvalues depend on $p$. 
\end{remark}

\begin{theorem} \label{thm:lnr}
A linearly parameterized $k$-channel system $\{A(p), B_i(p), C_i(p); k\}$ is structurally complete if and only if $\grank (A(p)+B(p)F(\tilde{p})C(p)) = n$ and $\forall \mathcal{S} \subset \mathbf{k}$, $\exists p \in {\rm I\!R}^q$ such that $\left\langle A \, | \, \mathscr{B}_{\mathcal{S}} \right\rangle$ is not a proper subspace of $[C_{\mathbf{k} - \mathcal{S}} \, | \, A]$.
\end{theorem}

\begin{remark} \label{rmk:lnr}
If $\grank (A(p)+B(p)F(\tilde{p})C(p)) < n$, the system has 0 in its fixed spectrum for all parameter values. This is equivalent to condition (\lowercase\expandafter{\romannumeral2}) in the Theorem of \cite{sezer1981structurally} by Proposition \ref{prp:mtrx-pencil} in Section \ref{sec:analyses}. While 0 is a fixed eigenvalue, it may or may not be a firmly fixed eigenvalue. On the other hand, if $\exists \mathcal{S} \subset \mathbf{k}$ such that $\forall p \in {\rm I\!R}^q$, $\left\langle A \, | \, \mathscr{B}_{\mathcal{S}} \right\rangle$ is a proper subspace of $[C_{\mathbf{k} - \mathcal{S}} \, | \, A]$, it follows from Remark \ref{rmk:poly} that the system has at least one firmly fixed eigenvalue for each $p \in {\rm I\!R}^q$. 
\end{remark}

\begin{theorem} \label{thm:binary}
Let $\{A(p), B_i(p), C_i(p); k\}$ be a linearly parameterized $k$-channel system which satisfies the binary assumption. The following statements are equivalent. \\
(\lowercase\expandafter{\romannumeral1}) The system $\{A(p), B_i(p), C_i(p); k\}$ is structurally complete. \\
(\lowercase\expandafter{\romannumeral2}) $\grank (A(p)+B(p)F(\tilde{p})C(p)) = n$ and there exist no subset $\mathcal{S} \subset \mathbf{k}$ and no permutation matrix $\Pi$ such that
\begin{align} \label{eqn:ABC-partition}
& \Pi A(p) \Pi^{-1} = 
\begin{bmatrix}
A_{11} & \mathbf{0} & \mathbf{0} \\
A_{21} & A_{22} & \mathbf{0} \\
A_{31} & A_{32} & A_{33}
\end{bmatrix}, \quad
\Pi B_{\mathcal{S}}(p) = 
\begin{bmatrix}
\mathbf{0} \\
\mathbf{0} \\
B_3
\end{bmatrix} \nonumber \\
& C_{\mathbf{k} - \mathcal{S}}(p) \Pi^{-1} = 
\begin{bmatrix}
C_1 & \mathbf{0} & \mathbf{0}
\end{bmatrix}
\end{align}
where $A_{11}$ is an $n_1 \times n_1$ block, $A_{33}$ is an $n_3 \times n_3$ block, $n_1 + n_3 < n$. \\
(\lowercase\expandafter{\romannumeral3}) The feedback graph $\mathbb{G}_F$ has an unbalanced similarity class of multi-colored cycle subgraphs, and each strongly connected component of $\mathbb{G}_F$ consists of either an input or output vertex or at least one input vertex, one state vertex, and one output vertex.
\end{theorem}

\begin{remark} \label{rmk:binary-algbr-graph}
Note that the algebraic condition in Theorem \ref{thm:binary} is the same as the condition for systems satisfying the unitary assumption \cite{sezer1981structurally}. This pattern is also observed in the structural controllability problem, where the algebraic condition for linearly parameterized systems satisfying the binary assumption is the same as the condition for systems satisfying the unitary assumption \cite{liu2019graphical, lin1974structural}. Obviously, the graphical condition does not follow this pattern. Condition (\lowercase\expandafter{\romannumeral3}) in Theorem \ref{thm:binary} reduces to Theorem 4 of \cite{pichai1984graph} provided that the system satisfies the unitary assumption. 
\end{remark}

\begin{remark} \label{rmk:binary-firm-fix-eigval}
If there exist $\mathcal{S} \subset \mathbf{k}$ and a permutation matrix $\Pi$ such that (\ref{eqn:ABC-partition}) holds, it follows from Remark \ref{rmk:poly} that the spectrum of $A_{22}$ belongs to the firmly fixed spectrum of the system. 
\end{remark}

\begin{remark} \label{rmk:binary-genericity}
The three theorems above imply that if one allows all entries of the system coefficient matrices to vary independently rather than imposes a specific type of parameterization on them, having no fixed spectrum is a generic property of a multi-channel linear system. 
\end{remark}


As an example of Theorem \ref{thm:binary}, the 2-channel system given in (\ref{eqn:graph-expl}) is structurally complete because its feedback graph in Figure \ref{fig:Graph-F} satisfies condition (\lowercase\expandafter{\romannumeral3}).

\section{Analyses} \label{sec:analyses}

This section focuses on the analyses and proofs of Theorem \ref{thm:poly}, Theorem \ref{thm:lnr}, and Theorem \ref{thm:binary}.

\subsection{Proof of Theorem \ref{thm:poly}}

A test for checking whether $\lambda \in \sigma(A)$ is a fixed eigenvalue of (\ref{eqn:syst}) is cited as follows, which is a direct result of Theorem 4.1 in \cite{anderson1981algebraic}.

\begin{proposition} \textnormal{\cite{anderson1981algebraic}}  \label{prp:mtrx-pencil}
A $k$-channel linear system $\{A, B_i, C_i; k\}$ has $\lambda \in \sigma(A)$ in its fixed spectrum if and only if $\exists \mathcal{S} \subset \mathbf{k}$ such that
\begin{equation*}
\rank 
\begin{bmatrix}
\lambda I_n - A & B_{\mathcal{S}} \\
C_{\mathbf{k}-\mathcal{S}} & \mathbf{0}
\end{bmatrix}
< n
\end{equation*}
\end{proposition}

Proposition \ref{prp:mtrx-pencil} reveals that whether a multi-channel linear system has a fixed spectrum is in fact a combinatorial problem involving all of its complementary subsystems \cite{corfmat1976decentralized} of the form $(C_{\mathbf{k}-\mathcal{S}}, A, B_{\mathcal{S}})$, $\mathcal{S} \subset \mathbf{k}$. For a proof of Proposition \ref{prp:mtrx-pencil}, please refer to \cite{anderson1981algebraic} or \cite{liu2019matroid}.

Two lemmas are needed to prove Theorem \ref{thm:poly}. More specifically, Lemma \ref{lem:pencil-rank} draws a connection between (\ref{eqn:poly-partition}) and (\ref{eqn:poly-fixed-eigval}), and Lemma \ref{lem:partition} shows how generic (\ref{eqn:poly-partition}) is in the parameter space.

\begin{lemma} \label{lem:pencil-rank}
Let matrices $A \in {\rm l \hspace{-0.45em} C}^{n \times n}$, $B \in {\rm l \hspace{-0.45em} C}^{n \times m}$, and $C \in {\rm l \hspace{-0.45em} C}^{l \times n}$. Then $
\rank 
\begin{bmatrix}
A & B \\
C & \mathbf{0}
\end{bmatrix}
\geq n
$
if and only if there exist matrices $E \in {\rm l \hspace{-0.45em} C}^{m \times n}$ and $K \in {\rm l \hspace{-0.45em} C}^{n \times l}$ such that $\rank (A + BE + KC) = n$.
\end{lemma}

\noindent \textbf{Proof of Lemma \ref{lem:pencil-rank}:} (Sufficiency) Suppose $
\rank 
\begin{bmatrix}
A & B \\
C & \mathbf{0}
\end{bmatrix}
< n
$, then $
\rank 
\begin{bmatrix}
A+BE+KC & B \\
C & \mathbf{0}
\end{bmatrix}
< n
$
for any matrices $E \in {\rm l \hspace{-0.45em} C}^{m \times n}$ and $K \in {\rm l \hspace{-0.45em} C}^{n \times l}$, as the rank of a matrix remains unchanged under elementary row and column operations. Thus $\rank (A + BE + KC) < n$ for any $E \in {\rm l \hspace{-0.45em} C}^{m \times n}$ and $K \in {\rm l \hspace{-0.45em} C}^{n \times l}$.

(Necessity) If $
\rank 
\begin{bmatrix}
A & B \\
C & \mathbf{0}
\end{bmatrix}
\geq n
$, by elementary column operations, $\exists E \in {\rm l \hspace{-0.45em} C}^{m \times n}$ such that $
\rank 
\begin{bmatrix}
A + BE \\
C 
\end{bmatrix}
= n
$. Similarly, by elementary row operations, $\exists K \in {\rm l \hspace{-0.45em} C}^{n \times l}$ such that $\rank (A + BE + KC) = n$. \hfill $\qed$

\begin{lemma} \label{lem:partition}
Let $\{A(p), B_i(p), C_i(p); k\}$ be a polynomially parameterized $k$-channel system. Given $\mathcal{S} \subset \mathbf{k}$, if $\exists p \in {\rm I\!R}^q$ such that (\ref{eqn:poly-partition}) holds, then (\ref{eqn:poly-partition}) holds for almost all $p \in {\rm I\!R}^q$. 
\end{lemma}

\noindent \textbf{Proof of Lemma \ref{lem:partition}:} Let $\mathcal{P}^*$ be the set of $p \in {\rm I\!R}^q$ for which (\ref{eqn:poly-partition}) holds. Let
\begin{equation} \label{eqn:M}
M(p, \lambda) \triangleq 
\begin{bmatrix}
\lambda I_n -A(p) & B_{\mathcal{S}}(p) \\
C_{\mathbf{k}-\mathcal{S}}(p) & 0
\end{bmatrix}
\end{equation}
and let $\mathbf{M}_{\rm I\!R}$ denote the real algebraic variety of matrices of the same size as $M$. Let $\mathbf{M}_{\rm I\!R}^{<n}$ be the closed subvariety of $\mathbf{M}_{\rm I\!R}$ defined by polynomial equations that the determinants of all $n \times n$ submatrices of $M$ are 0. Then $\mathbf{M}_{\rm I\!R}^{<n} ({\rm l \hspace{-0.45em} C})$ consists of exactly those complex matrices $M$ of rank less than $n$.

Now we identify $\Spec {\rm I\!R}[P_1, P_2, \dots, P_q]$ with the real affine space $\mathbf{A}_{{\rm I\!R}}^q$ of dimension $q$, and $\Spec {\rm I\!R}[P_1, P_2, \dots, P_q, \Lambda]$ with the real affine space $\mathbf{A}_{{\rm I\!R}}^{q+1}$ of dimension $q+1$. Then there is a morphism $\pi: \mathbf{A}_{{\rm I\!R}}^{q+1} \to \mathbf{A}_{{\rm I\!R}}^q$ by projecting to the first $q$ coordinates; and the matrix $M$ in (\ref{eqn:M}) defines a morphism $\mu: \mathbf{A}_{{\rm I\!R}}^{q+1} \to \mathbf{M}_{\rm I\!R}$.

It is claimed that $\pi (\mu^{-1} \mathbf{M}^{<n}_{{\rm I\!R}})$ is (Zariski) closed in $\mathbf{A}_{{\rm I\!R}}^q$. Assuming this claim, then $\mathcal{P}^*$ is the complement of $\pi (\mu^{-1} \mathbf{M}^{<n}_{{\rm I\!R}})({\rm I\!R})$ in $\mathbf{A}_{{\rm I\!R}}^q({\rm I\!R}) = {\rm I\!R}^q$. As $\mathcal{P}^*$ is not empty, $\pi (\mu^{-1} \mathbf{M}^{<n}_{{\rm I\!R}})$ is not the entire affine space $\mathbf{A}_{{\rm I\!R}}^q$, which implies that $\pi (\mu^{-1} \mathbf{M}^{<n}_{{\rm I\!R}})({\rm I\!R})$ has Lebesgue measure zero.

Next the claim will be proved. Let $\Phi \triangleq \det (\Lambda I_n - A)$, which is an element in ${\rm I\!R}[P_1, P_2, \dots, P_q, \Lambda]$. It is not hard to see that if $(p_1, p_2, \dots, p_q, \lambda)$ is not a root of $\Phi$, $M(p, \lambda)$ has rank at least $n$. Therefore, $\mu^{-1} \mathbf{M}^{<n}_{{\rm I\!R}}$ is contained in the closed subvariety $\mathbf{S} \triangleq \Spec {\rm I\!R}[P_1, P_2, \dots, P_q, \Lambda]/(\Phi)$ of $\mathbf{A}_{{\rm I\!R}}^{q+1}$. Let $\pi_{\mathbf{S}}: \mathbf{S} \to \mathbf{A}_{{\rm I\!R}}^q$ denote the restriction of $\pi$ to $\mathbf{S}$. Since $\Phi$ has leading coefficient 1 as a polynomial in $\Lambda$, the morphism $\pi_{\mathbf{S}}$ is finite. By Exercise 4.1 in \cite{algebraic1977geometry}, $\pi_{\mathbf{S}}$ is proper. Thus $\pi_{\mathbf{S}} (\mu^{-1} \mathbf{M}^{<n}_{{\rm I\!R}})$ is (Zariski) closed as $\mu^{-1} \mathbf{M}^{<n}_{{\rm I\!R}}$ is closed. \hfill $\qed$

\noindent \textbf{Proof of Theorem \ref{thm:poly}:} By Lemma \ref{lem:pencil-rank}, (\ref{eqn:poly-partition}) holds if and only if for every $\lambda \in \sigma(A)$, there exist matrices $E$ and $K$ of appropriate sizes such that 
\begin{equation*}
\rank \left( \lambda I_n - A - B_{\mathcal{S}}E - KC_{\mathbf{k}-\mathcal{S}} \right) = n
\end{equation*} 
i.e., $\lambda$ is not an eigenvalue of $A + B_{\mathcal{S}}E + KC_{\mathbf{k}-\mathcal{S}}$. This establishes the equivalence between (\ref{eqn:poly-partition}) and (\ref{eqn:poly-fixed-eigval}), so it suffices to prove the necessary and sufficient condition involving (\ref{eqn:poly-partition}). By the definition of structural completeness and Proposition \ref{prp:mtrx-pencil}, the system $\{A(p), B_i(p), C_i(p); k\}$ is structurally complete if and only if $\exists p \in {\rm I\!R}^q$ such that $\forall \mathcal{S} \subset \mathbf{k}$, (\ref{eqn:poly-partition}) holds. So the necessity part of Theorem \ref{thm:poly} is obvious. To see why the inverse is true, suppose $\forall \mathcal{S} \subset \mathbf{k}$, $\exists p \in {\rm I\!R}^q$ such that (\ref{eqn:poly-partition}) holds. By Lemma \ref{lem:partition}, $\forall \mathcal{S} \subset \mathbf{k}$, (\ref{eqn:poly-partition}) holds for almost all $p \in {\rm I\!R}^q$. As there are only finite choices of $\mathcal{S}$, $\exists p \in {\rm I\!R}^q$ such that $\forall \mathcal{S} \subset \mathbf{k}$, (\ref{eqn:poly-partition}) holds. The proof for sufficiency is complete. \hfill $\qed$

\subsection{Proof of Theorem \ref{thm:lnr}}

In the same spirit of the linear parameterization defined by (\ref{eqn:lnr-param-def}), a matrix pair $(A(p), B(p))$ is linearly parameterized if it is of the form
\begin{equation} \label{eqn:lnr-param-pair}
\begin{bmatrix}
A_{n \times n}(p) \enspace B_{n \times m}(p)
\end{bmatrix}
=
\sum_{i \in \mathbf{q}} g_i p_i h_i
\end{equation}
where $g_i \in {\rm I\!R}^n$ and $h_i \in {\rm I\!R}^{1 \times (n+m)}$. A linearly parameterized pair $(A(p), B(p))$ is said to be \emph{structurally controllable} if there exists a parameter vector $p \in {\rm I\!R}^q$ for which $(A, B)$ is controllable. It is not hard to see that structural controllability implies controllability for almost every value of $p$. Lemma \ref{lem:struc-contr} and Corollary \ref{cor:struc-contr} below provide necessary and sufficient conditions for the structural controllability of a linearly parameterized matrix pair. It will be shown that these conditions are equivalent to the one in Proposition 3 of \cite{liu2019graphical}.

\begin{lemma} \label{lem:struc-contr}
A linearly parameterized matrix pair $(A(p), B(p))$ given by (\ref{eqn:lnr-param-pair}) is structurally controllable if and only if $\grank [A(p) \enspace B(p)]=n$ and every parameter in $(A(p), B(p))$ appears in the matrix 
\begin{equation} \label{eqn:long-matrix1}
\begin{bmatrix}
B(p) \enspace A(p)B(p) \enspace \dots \enspace A^i(p)B(p)
\end{bmatrix}
\end{equation}
for some nonnegative integer $i$.
\end{lemma}

\noindent \textbf{Proof of Lemma \ref{lem:struc-contr}:} By the definition of the transfer graph $\mathbb{T}$ in \cite{liu2019graphical}, it is claimed that there is a directed path of length $j+1 >0$ in $\mathbb{T}$ from vertex 0 to vertex $\alpha \in \mathbf{q}$ if and only if parameter $p_{\alpha}$ appears in $A^j(p)B(p)$. So the transfer graph $\mathbb{T}$ has a spanning tree rooted at vertex 0 if and only if every parameter in $(A(p), B(p))$ appears in the matrix (\ref{eqn:long-matrix1}) for some $i \geq 0$. By Proposition 3 and Lemma 2 in \cite{liu2019graphical}, Lemma \ref{lem:struc-contr} is true.

Next a slightly stronger statement than the claim will be proved. The statement is that there is a directed path of length $j+1 >0$ in $\mathbb{T}$ from vertex 0 to vertex $\alpha \in \mathbf{q}$ if and only if a nonzero scalar multiple of $g_{\alpha} p_{\alpha}$ is contained in a column of $A^j(p)B(p)$. It will be proved by induction on the length of a path from vertex 0 in $\mathbb{T}$. By the definition of $\mathbb{T}$, there is an arc from vertex 0 to vertex $\alpha$ if and only if $h_{\alpha 2} \neq 0$ \cite{liu2019graphical}. Suppose the $e$th entry of $h_{\alpha 2}$ is nonzero for some $e \in \{1, 2, \dots, m\}$, then by equation (2) in \cite{liu2019graphical}, the $e$th column of $B(p)$ contains a nonzero scalar multiple of $g_{\alpha} p_{\alpha}$. So the statement is true for $j=0$. Now suppose the statement holds for $j=i \geq 0$ and consider the case when $j=i+1$. As there is a directed path of length $i+2$ in $\mathbb{T}$ from vertex 0 to vertex $\alpha$, there exists a vertex $\beta \in \mathbf{q}$ such that $\mathbb{T}$ has a directed path of length $i+1$ from vertex 0 to vertex $\beta$ and has an arc from vertex $\beta$ to vertex $\alpha$. By the induction hypothesis, a column of $A^i(p)B(p)$ contains a nonzero scalar multiple of $g_{\beta} p_{\beta}$. By the definition of $\mathbb{T}$, $h_{\alpha 1} g_{\beta} \neq 0$. In particular, $h_{\alpha 1} \neq 0$ means that $A(p)$ contains $g_{\alpha} p_{\alpha} h_{\alpha 1}$ \cite{liu2019graphical}. As $h_{\alpha 1} g_{\beta} \neq 0$, a nonzero scalar multiple of $g_{\alpha} p_{\alpha}$ is contained in a column of $A(p) \cdot A^i(p)B(p) = A^{i+1}(p)B(p)$. Thus the statement is true for $j=i+1$. This completes the proof of the statement. \hfill $\qed$

\begin{corollary} \label{cor:struc-contr}
A linearly parameterized matrix pair $(A(p), B(p))$ given by (\ref{eqn:lnr-param-pair}) is structurally controllable if and only if $\grank [A(p) \enspace B(p)]=n$ and every parameter in $(A(p), B(p))$ appears in the matrix 
\begin{equation} \label{eqn:long-matrix2}
\begin{bmatrix}
B(p) \enspace A(p)B(p) \enspace \dots \enspace A^n(p)B(p)
\end{bmatrix}
\end{equation}
\end{corollary}

\noindent \textbf{Proof of Corollary \ref{cor:struc-contr}:} Let 
\begin{equation*}
M(p) = [B(p) \enspace A(p)B(p) \enspace \dots \enspace A^{n-1}(p)B(p)]
\end{equation*}
If $(A(p), B(p))$ is structurally controllable, $\grank M(p)=n$. Then all parameters in $A(p)$ will appear in $A(p)M(p)$. So all parameters in $(A(p), B(p))$ will appear in $[B(p) \enspace A(p)M(p)]$, which is matrix (\ref{eqn:long-matrix2}). As $\grank M(p)=n$ and $\Image M(p) \subset \Image [A(p) \enspace B(p)]$, $\grank [A(p) \enspace B(p)]=n$. \hfill $\qed$

The result below from Remark 1 in \cite{corfmat1976control} is also needed for the following proof of Theorem \ref{thm:lnr}.

\begin{proposition} \textnormal{\cite{corfmat1976control}} \label{prp:remnant}
Let $(C, A, B)$ be a standard single-channel linear system. If $C(\lambda I - A)^{-1}B \neq 0$, the uncontrollable polynomial of $(A+KC, B)$ equals the remnant polynomial \cite{corfmat1976control} of $(C, A, B)$ whenever $K$ is selected so that the dimension of $\left\langle A+KC \, | \, \mathscr{B} \right\rangle$ is as large as possible.
\end{proposition}

\noindent \textbf{Proof of Theorem \ref{thm:lnr}:} (Necessity) If $\grank (A(p)+B(p)F(\tilde{p})C(p)) < n$, matrix $A+BFC$ has a fixed eigenvalue of 0 for all parameter values $p$ and $\tilde{p}$. Thus the system $\{A(p), B_i(p), C_i(p); k\}$ is structurally incomplete. If $\exists \mathcal{S} \subset \mathbf{k}$ such that $\forall p \in {\rm I\!R}^q$, $\left\langle A \, | \, \mathscr{B}_{\mathcal{S}} \right\rangle$ is a proper subspace of $[C_{\mathbf{k} - \mathcal{S}} \, | \, A]$, then for each fixed $p \in {\rm I\!R}^q$, there exists a nonsingular matrix $T$ such that 
\begin{align*}
& T A T^{-1} = 
\begin{bmatrix}
A_{11} & \mathbf{0} & \mathbf{0} \\
A_{21} & A_{22} & \mathbf{0} \\
A_{31} & A_{32} & A_{33}
\end{bmatrix}, \quad
T B_{\mathcal{S}} = 
\begin{bmatrix}
\mathbf{0} \\
\mathbf{0} \\
B_3
\end{bmatrix} \\
& C_{\mathbf{k} - \mathcal{S}} T^{-1} = 
\begin{bmatrix}
C_1 & \mathbf{0} & \mathbf{0}
\end{bmatrix}
\end{align*}
where $A_{11}$ is an $n_1 \times n_1$ block, $A_{33}$ is an $n_3 \times n_3$ block, $n_1 + n_3 < n$. So $\sigma (A_{22})$ is in the fixed spectrum of system $\{A, B_i, C_i; k\}$. As this is true for each fixed $p \in {\rm I\!R}^q$, the linearly parameterized system $\{A(p), B_i(p), C_i(p); k\}$ is structurally incomplete.

(Sufficiency) If a linearly parameterized system $\{A(p), B_i(p), C_i(p); k\}$ is structurally incomplete, by Theorem \ref{thm:poly}, $\exists \mathcal{S} \subset \mathbf{k}$ such that $\forall p \in {\rm I\!R}^q$, 
\begin{equation} \label{eqn:poly-partition-neg}
\rank 
\begin{bmatrix}
\lambda I_n -A & B_{\mathcal{S}} \\
C_{\mathbf{k}-\mathcal{S}} & \mathbf{0}
\end{bmatrix}
< n \quad \text{for some} \enspace \lambda \in \sigma(A)
\end{equation}
or equivalently, 
\begin{equation} \label{eqn:poly-fixed-eigval-neg}
\bigcap_{\substack{E_i \in {\rm I\!R}^{m_i \times n}, \, i \in \mathcal{S} \\ K_j \in {\rm I\!R}^{n \times l_j}, \, j \in \mathbf{k}-\mathcal{S}}} \sigma \left( A + \sum_{i \in \mathcal{S}} B_i E_i + \sum_{j \in \mathbf{k}-\mathcal{S}} K_j C_j \right) \neq \emptyset
\end{equation} 
Depending on the value of the transfer function $C_{\mathbf{k} - \mathcal{S}}(p) (\lambda I - A(p))^{-1} B_{\mathcal{S}}(p)$, two cases are discussed as follows.

Case 1: $C_{\mathbf{k} - \mathcal{S}}(p) (\lambda I - A(p))^{-1} B_{\mathcal{S}}(p) \equiv 0$. This is equivalent to the condition that $\forall p \in {\rm I\!R}^q$, $\left\langle A \, | \, \mathscr{B}_{\mathcal{S}} \right\rangle \subset [C_{\mathbf{k} - \mathcal{S}} \, | \, A]$. If $\exists p \in {\rm I\!R}^q$ such that $\left\langle A \, | \, \mathscr{B}_{\mathcal{S}} \right\rangle = [C_{\mathbf{k} - \mathcal{S}} \, | \, A]$, then the spectrum of $A+B_{\mathcal{S}}E_{\mathcal{S}}+E_{\mathbf{k} - \mathcal{S}}C_{\mathbf{k} - \mathcal{S}}$ can be freely assigned with suitable matrices $E_{\mathcal{S}}$ and $E_{\mathbf{k} - \mathcal{S}}$, which violates (\ref{eqn:poly-fixed-eigval-neg}). So $\mathcal{S} \subset \mathbf{k}$ is such that $\forall p \in {\rm I\!R}^q$, $\left\langle A \, | \, \mathscr{B}_{\mathcal{S}} \right\rangle$ is a proper subspace of $[C_{\mathbf{k} - \mathcal{S}} \, | \, A]$.

Case 2: $C_{\mathbf{k} - \mathcal{S}}(p) (\lambda I - A(p))^{-1} B_{\mathcal{S}}(p) \neq 0$. That is, for almost all $p \in {\rm I\!R}^q$, $C_{\mathbf{k} - \mathcal{S}} (\lambda I - A)^{-1} B_{\mathcal{S}} \neq 0$. By Corollary 4 in \cite{corfmat1976control}, (\ref{eqn:poly-partition-neg}) implies that for almost all $p \in {\rm I\!R}^q$, the triple $(C_{\mathbf{k} - \mathcal{S}}, A, B_{\mathcal{S}})$ is not complete. That is, for almost all $p \in {\rm I\!R}^q$, the remnant polynomial of $(C_{\mathbf{k} - \mathcal{S}}, A, B_{\mathcal{S}})$ is not 1. By Proposition \ref{prp:remnant}, it means that for almost all $p \in {\rm I\!R}^q$, $(A+KC_{\mathbf{k} - \mathcal{S}}, B_{\mathcal{S}})$ is not controllable for any matrix $K$ of appropriate size. This is equivalent to the statement that for any given matrix $K$, $(A+KC_{\mathbf{k} - \mathcal{S}}, B_{\mathcal{S}})$ is not controllable for almost all $p \in {\rm I\!R}^q$. Note that for every fixed matrix $K$, $(A(p) + KC_{\mathbf{k} - \mathcal{S}}(p), B_{\mathcal{S}}(p))$ is a linearly parameterized matrix pair. Thus, for any fixed matrix $K$, the pair $(A(p) + KC_{\mathbf{k} - \mathcal{S}}(p), B_{\mathcal{S}}(p))$ is not structurally controllable. Since $C_{\mathbf{k} - \mathcal{S}}(p) (\lambda I - A(p))^{-1} B_{\mathcal{S}}(p) \neq 0$, let $i \in \{0, 1, \dots, n-1\}$ be the smallest integer for which $C_{\mathbf{k} - \mathcal{S}}(p) A^i(p) B_{\mathcal{S}}(p) \neq 0$, then 
\begin{multline*}
\left( A(p) + KC_{\mathbf{k} - \mathcal{S}}(p) \right)^{i+1} B_{\mathcal{S}}(p) = A^{i+1}(p) B_{\mathcal{S}}(p) + \\
KC_{\mathbf{k} - \mathcal{S}}(p) A^i(p) B_{\mathcal{S}}(p)
\end{multline*}
For almost all matrix $K$, 
\begin{equation*}
A^{i+1}(p)B_{\mathcal{S}}(p) + KC_{\mathbf{k} - \mathcal{S}}(p) A^i(p) B_{\mathcal{S}}(p)
\end{equation*} 
has a column in which every entry is nonzero. So for almost all matrix $K$, every parameter in $A(p) + KC_{\mathbf{k} - \mathcal{S}}(p)$ appears in $(A(p) + KC_{\mathbf{k} - \mathcal{S}}(p))^{i+2} B_{\mathcal{S}}(p)$, thus every parameter in $(A(p) + KC_{\mathbf{k} - \mathcal{S}}(p), B_{\mathcal{S}}(p))$ appears in the matrix 
\begin{multline*} 
\left[
B_{\mathcal{S}}(p) \quad (A(p)+KC_{\mathbf{k} - \mathcal{S}}(p))B_{\mathcal{S}}(p) \quad \dots 
\right. \\
\left. 
(A(p)+KC_{\mathbf{k} - \mathcal{S}}(p))^{i+2} B_{\mathcal{S}}(p)
\right]
\end{multline*}
As the pair $(A(p) + KC_{\mathbf{k} - \mathcal{S}}(p), B_{\mathcal{S}}(p))$ is not structurally controllable for any fixed matrix $K$, Lemma \ref{lem:struc-contr} implies that for almost all matrix $K$, $\grank [A(p)+KC_{\mathbf{k}-\mathcal{S}}(p) \enspace B_{\mathcal{S}}(p)]<n$. It follows immediately that for every fixed matrix $K$, $\grank [A(p)+KC_{\mathbf{k}-\mathcal{S}}(p) \enspace B_{\mathcal{S}}(p)]<n$. Therefore,  
\begin{equation*}
\grank
\begin{bmatrix}
A(p) & B_{\mathcal{S}}(p) \\
C_{\mathbf{k}-\mathcal{S}}(p) & \mathbf{0}
\end{bmatrix}
< n
\end{equation*}
By Proposition \ref{prp:mtrx-pencil}, the system $\{A, B_i, C_i; k\}$ has 0 in its fixed spectrum for all $p \in {\rm I\!R}^q$. That is, $\grank (A(p)+B(p)F(\tilde{p})C(p)) < n$. \hfill $\qed$

\subsection{Proof of Theorem \ref{thm:binary}}

In the same spirit of the binary assumption defined before, a linearly parameterized matrix pair $(A(p), B(p))$ given by (\ref{eqn:lnr-param-pair}) satisfies the binary assumption if all of the $g_i$ and $h_i$ appearing in (\ref{eqn:lnr-param-pair}) are binary vectors. Let $'$ denote transposition. Generalizing the standard notion of irreducibility, a matrix pair $(A, B)$ is said to be \emph{irreducible} if there is no permutation matrix $\Pi$ bringing $(A, B)$ into the form
\begin{equation*}
\Pi A \Pi^{-1} =
\begin{bmatrix}
A_{11} & \mathbf{0} \\
A_{21} & A_{22}
\end{bmatrix}, \quad
\Pi B =
\begin{bmatrix}
\mathbf{0} \\
B_2
\end{bmatrix}
\end{equation*}
where $A_{11}$ is an $n_1 \times n_1$ block, $0 < n_1 < n$.

The following result on the structural controllability of a linearly parameterized matrix pair satisfying the binary assumption is from Theorem 1 in \cite{liu2019graphical} and will be used to prove Theorem \ref{thm:binary} in this paper.

\begin{proposition} \textnormal{\cite{liu2019graphical}} \label{prp:struc-contr}
Let $(A_{n \times n}(p), B_{n \times m}(p))$ be a linearly parameterized matrix pair which satisfies the binary assumption. Then the pair $(A(p), B(p))$ is structurally controllable if and only if $\grank \left[ A(p) \enspace B(p) \right] = n$ and $(A(p), B(p))$ is irreducible. 
\end{proposition}

In addition to Proposition \ref{prp:struc-contr}, Lemma \ref{lem:grk-acc} facilitates the proof of the algebraic condition in Theorem \ref{thm:binary}, and Lemma \ref{lem:grk-graph} shows half of the graphical condition.

\begin{lemma} \label{lem:grk-acc}
Let $\{A(p), B_i(p), C_i(p); k\}$ be a linearly parameterized $k$-channel system which satisfies the binary assumption. If $\grank (A(p)+B(p)F(\tilde{p})C(p)) = n$ and there exist $\mathcal{S} \subset \mathbf{k}$ and a permutation matrix $\Pi$ such that
\begin{align} \label{eqn:permute}
& \Pi A(p) \Pi^{-1} = 
\begin{bmatrix}
A_{11} & \mathbf{0} \\
A_{21} & A_{22}
\end{bmatrix}, \quad
\Pi B_{\mathcal{S}}(p) = 
\begin{bmatrix}
\mathbf{0} \\
B_2
\end{bmatrix} \nonumber \\
& C_{\mathbf{k} - \mathcal{S}}(p) \Pi^{-1} = 
\begin{bmatrix}
C_1 & \mathbf{0}
\end{bmatrix}
\end{align}
where both pairs $(A_{22}, B_2)$ and $(A_{11}', C_1')$ are irreducible, then $\exists p \in {\rm I\!R}^q$ for which $\left\langle A \, | \, \mathscr{B}_{\mathcal{S}} \right\rangle = [C_{\mathbf{k} - \mathcal{S}} \, | \, A]$.
\end{lemma}

\noindent \textbf{Proof of Lemma \ref{lem:grk-acc}:} Let the size of matrix $C_{\mathbf{k} - \mathcal{S}}$ be $l_{\mathbf{k} - \mathcal{S}} \times n$ and let the size of matrix $B_{\mathcal{S}}$ be $n \times m_{\mathcal{S}}$. It is easy to see that 
\begin{multline} \label{eqn:permute-all}
\begin{bmatrix}
\Pi A(p) \Pi^{-1} & \Pi B_{\mathcal{S}}(p) \\
C_{\mathbf{k} - \mathcal{S}}(p) \Pi^{-1} & \mathbf{0}
\end{bmatrix}
= \\
\begin{bmatrix}
\Pi & \mathbf{0} \\
\mathbf{0} & I_{l_{\mathbf{k} - \mathcal{S}}}
\end{bmatrix}
\begin{bmatrix}
A(p) & B_{\mathcal{S}}(p) \\
C_{\mathbf{k} - \mathcal{S}}(p) & \mathbf{0}
\end{bmatrix}
\begin{bmatrix}
\Pi^{-1} & \mathbf{0} \\
\mathbf{0} & I_{m_{\mathcal{S}}}
\end{bmatrix}
\end{multline}
As $\{A(p), B_i(p), C_i(p); k\}$ is linearly parameterized and satisfies the binary assumption, 
\begin{equation} \label{eqn:binary-vector}
\begin{bmatrix}
A(p) & B_{\mathcal{S}}(p) \\
C_{\mathbf{k} - \mathcal{S}}(p) & \mathbf{0}
\end{bmatrix}
=
\sum_{i \in \mathbf{q}} \bar{g}_i p_i \bar{h}_i
\end{equation}
where $\bar{g}_i \in {\rm I\!R}^{n + l_{\mathbf{k} - \mathcal{S}}}$ and $\bar{h}_i \in {\rm I\!R}^{n + m_{\mathcal{S}}}$ are binary vectors for each $i \in \mathbf{q}$. Let 
\begin{equation} \label{eqn:change-vector}
\widehat{g}_i = 
\begin{bmatrix}
\Pi & \mathbf{0} \\
\mathbf{0} & I_{l_{\mathbf{k} - \mathcal{S}}}
\end{bmatrix}
\bar{g}_i
, \qquad
\widehat{h}_i = 
\bar{h}_i
\begin{bmatrix}
\Pi^{-1} & \mathbf{0} \\
\mathbf{0} & I_{m_{\mathcal{S}}}
\end{bmatrix}
\end{equation}
then $\widehat{g}_i \in {\rm I\!R}^{n + l_{\mathbf{k} - \mathcal{S}}}$ and $\widehat{h}_i \in {\rm I\!R}^{n + m_{\mathcal{S}}}$ are also binary vectors for each $i \in \mathbf{q}$. Combining (\ref{eqn:permute-all}), (\ref{eqn:binary-vector}), and (\ref{eqn:change-vector}), we get 
\begin{equation*}
\begin{bmatrix}
\Pi A(p) \Pi^{-1} & \Pi B_{\mathcal{S}}(p) \\
C_{\mathbf{k} - \mathcal{S}}(p) \Pi^{-1} & \mathbf{0}
\end{bmatrix}
=
\sum_{i \in \mathbf{q}} \widehat{g}_i p_i \widehat{h}_i
\end{equation*}
so both matrix pairs $(A_{22}(p), B_2(p))$ and $(A_{11}'(p), C_1'(p))$ are linearly parameterized and satisfy the binary assumption. Suppose $\mathcal{S} = \{i_1, i_2, \dots, i_s\}$ with $i_1 < i_2 < \dots < i_s$, then define $F_{\mathcal{S}} \triangleq \blkdiag \{F_{i_1}, F_{i_2}, \dots, F_{i_s}\}$. As 
\begin{equation*}
BFC = B_{\mathcal{S}}F_{\mathcal{S}}C_{\mathcal{S}}+B_{\mathbf{k} - \mathcal{S}}F_{\mathbf{k} - \mathcal{S}}C_{\mathbf{k} - \mathcal{S}}
\end{equation*}
the condition that $\grank \left( A(p)+B(p)F(\tilde{p})C(p) \right) = n$ implies  
\begin{multline} \label{eqn:permute-rank}
\grank \left( \Pi A(p) \Pi^{-1} + \Pi B_{\mathcal{S}}(p) F_{\mathcal{S}}(\tilde{p}) C_{\mathcal{S}}(p) \Pi^{-1} + \right. \\
\left. \Pi B_{\mathbf{k} - \mathcal{S}}(p) F_{\mathbf{k} - \mathcal{S}}(\tilde{p}) C_{\mathbf{k} - \mathcal{S}}(p) \Pi^{-1} \right) = n
\end{multline}
Suppose $A_{11}$ in (\ref{eqn:permute}) is an $n_1 \times n_1$ block, where $0 \leq n_1 \leq n$. Equations (\ref{eqn:permute}) and (\ref{eqn:permute-rank}) suggest that $\grank [A_{22}(p) \enspace B_2(p)] = n - n_1$ and $\grank [A_{11}'(p) \enspace C_1'(p)] = n_1$. Because matrix pairs $(A_{22}(p), B_2(p))$ and $(A_{11}'(p), C_1'(p))$ are both irreducible, by Proposition \ref{prp:struc-contr}, both pairs are structurally controllable. Therefore, $\exists p \in {\rm I\!R}^q$ for which $\left\langle A_{22} \, | \, \mathscr{B}_2 \right\rangle = [C_1 \, | \, A_{11}]$, which implies that $\left\langle \Pi A \Pi^{-1} \, | \, \Pi \mathscr{B}_{\mathcal{S}} \right\rangle = [C_{\mathbf{k} - \mathcal{S}} \Pi^{-1} \, | \, \Pi A \Pi^{-1}]$ and thus $\left\langle A \, | \, \mathscr{B}_{\mathcal{S}} \right\rangle = [C_{\mathbf{k} - \mathcal{S}} \, | \, A]$. \hfill $\qed$

\begin{lemma} \label{lem:grk-graph}
Let $\{A(p), B_i(p), C_i(p); k\}$ be a linearly parameterized $k$-channel system which satisfies the binary assumption. Then $\grank (A(p)+B(p)F(\tilde{p})C(p)) = n$ if and only if the feedback graph $\mathbb{G}_F$ has an unbalanced similarity class of multi-colored cycle subgraphs.
\end{lemma}

\noindent \textbf{Proof of Lemma \ref{lem:grk-graph}:} The matrix $A(p)+B(p)F(\tilde{p})C(p)$ has full generic rank if and only if its determinant is nonzero. The following proof will show that the determinant is nonzero if and only if the feedback graph $\mathbb{G}_F$ has an unbalanced similarity class of multi-colored cycle subgraphs. The proof takes three steps: First, the kind of subgraph of $\mathbb{G}_F$ that corresponds to a term in the determinant will be characterized. Second, some of those subgraphs that correspond to terms not appearing in the final expression of the determinant will be ruled out. Third, the graphical property for the sign of a term in the determinant will be specified. After these three steps, it can be determined purely from $\mathbb{G}_F$ that whether the determinant of $A(p)+B(p)F(\tilde{p})C(p)$ is identically zero.

The linear parameterization and the binary assumption imply that a nonzero entry of $A(p)+B(p)F(\tilde{p})C(p)$ is either a single parameter $p_r$ or a product of three parameters $p_r \tilde{p}_{\tilde{s}} p_t$, for some $r, t \in \mathbf{q}$, $r \neq t$, $\tilde{s} \in \tilde{\mathbf{q}} \triangleq \{1, 2, \dots, \tilde{q}\}$, or a sum of finitely many of them. A term in the determinant of $A(p)+B(p)F(\tilde{p})C(p)$ is a signed product of $n$ entries taken from different rows and columns. If any of the $n$ entries is a sum of multiple summands, the term in the determinant can be written as a sum of multiple subterms. So without loss of generality, assume each term in the determinant is a signed product of $n$ factors, each of which is either $p_r$ or $p_r \tilde{p}_{\tilde{s}} p_t$ for some $r, t \in \mathbf{q}$, $r \neq t$, $\tilde{s} \in \tilde{\mathbf{q}}$. A factor $p_r$ from the $ij$th entry of $A(p)+B(p)F(\tilde{p})C(p)$ corresponds to an arc $(x_j, x_i)_r$ of color $r$ in $\mathbb{G}_F$ from state vertex $x_j$ to state vertex $x_i$. A factor $p_r \tilde{p}_{\tilde{s}} p_t$ from the $ij$th entry of $A(p)+B(p)F(\tilde{p})C(p)$ corresponds to three arcs of distinct colors in $\mathbb{G}_F$: an arc $(x_j, y_d)_t$ from state vertex $x_j$ to an output vertex, an arc $(y_d, u_w)_{\tilde{s}}$ from the output vertex to an input vertex, and an arc $(u_w, x_i)_r$ from the input vertex to state vertex $x_i$, which together form a directed path from $x_j$ to $x_i$. As no two factors in a term of the determinant are taken from the same row or the same column of $A(p)+B(p)F(\tilde{p})C(p)$, each term in the determinant corresponds to the union of a finite number of cycle graphs with all state vertices contained in the union graph and each state vertex having exactly one incoming arc and one outgoing arc. For ease of reference, such a union graph is called a \emph{term subgraph} of $\mathbb{G}_F$.

If any output vertex in a term subgraph of $\mathbb{G}_F$ has two incoming arcs or two outgoing arcs, then in the corresponding term $z_1$ there are two factors taken from the matrix $B_i F_i^{:j} C_i^j$, where $F_i^{:j}$ is the $j$th column of $F_i$ and $C_i^j$ is the $j$th row of $C_i$, for some $i \in \mathbf{k}$, $j \in \{1, 2, \dots, l_i\}$. Because $\rank (B_i F_i^{:j} C_i^j) = 1$, in the same determinant there must be another term $z_2$, which is the product of the same $n$ factors as in term $z_1$ but with the opposite sign. So terms $z_1$ and $z_2$ cancel each other out and both of them do not appear in the final expression of the determinant. As the goal is to determine whether the determinant is identically zero, we can safely ignore those term subgraphs in which an output vertex has more than one incoming arc or more than one outgoing arc. Similarly, those term subgraphs in which an input vertex has more than one incoming arc or more than one outgoing arc can be safely ignored. Thus, we only care about those term subgraphs of $\mathbb{G}_F$, in which each vertex has exactly one incoming arc and one outgoing arc, i.e., those term subgraphs which are the disjoint unions of cycle graphs, because they correspond to terms that may appear in the final expression of the determinant of $A(p)+B(p)F(\tilde{p})C(p)$.

Next, it will be shown that the terms appearing in the final expression of the determinant correspond to the term subgraphs of $\mathbb{G}_F$, in which each arc is of a different color. In other words, only multi-colored cycle subgraphs of $\mathbb{G}_F$ matter. Two cases are considered as follows. Case 1: If a term $z_1$ has a parameter $p_r$ raised to the power of 2 for some $r \in \mathbf{q}$, then two factors in $z_1$ contain $p_r$. But linear parameterization of the system $\{A(p), B_i(p), C_i(p); k\}$ implies that $\forall r \in \mathbf{q}$, parameter $p_r$ appears in the matrix $A(p)+B(p)F(\tilde{p})C(p)$ in a rank-one fashion, i.e., 
\begin{multline*}
A(p)+B(p)F(\tilde{p})C(p) = \sum_{r \in \mathbf{q}_1} g_{r1} p_r h_{r3}(p, \tilde{p}) + \\
\sum_{r \in \mathbf{q}-\mathbf{q}_1} g_{r3}(p, \tilde{p}) p_r h_{r1}
\end{multline*} 
where $\mathbf{q}_1 \triangleq \{r \in \mathbf{q} \, | \, p_r \enspace \text{appears in} \enspace B(p)\}$, $g_{r1} \in {\rm I\!R}^n$ is the first $n$ entries of $g_r$ given by (\ref{eqn:lnr-param-def}), $h_{r3}(p, \tilde{p})$ is a parameterized row vector of size $n$, $g_{r3}(p, \tilde{p})$ is a parameterized column vector of size $n$, and $h_{r1} \in {\rm I\!R}^{1 \times n}$ is the first $n$ entries of $h_r$ given by (\ref{eqn:lnr-param-def}). So in the same determinant there must be another term $z_2$, which is the product of the same $n$ factors as in term $z_1$ but with the opposite sign. Thus, terms $z_1$ and $z_2$ cancel each other out and both of them do not appear in the final expression of the determinant. Case 2: If a term $z_1$ has a parameter $\tilde{p}_{\tilde{s}}$ raised to the power of 2 for some $\tilde{s} \in \tilde{\mathbf{q}}$, two factors in $z_1$ contain $\tilde{p}_{\tilde{s}}$. Suppose $\tilde{p}_{\tilde{s}}$ is the $ij$th entry of $F_t$ for some $t \in \mathbf{k}$, then the two factors in $z_1$ are taken from the rank-one matrix $B_t^i \tilde{p}_{\tilde{s}} C_t^j$, where $B_t^i$ is the $i$th column of $B_t$ and $C_t^j$ is the $j$th row of $C_t$. So in the same determinant there must be another term $z_2$, which is the product of the same $n$ factors as in term $z_1$ but with the opposite sign. Thus, terms $z_1$ and $z_2$ cancel each other out and both of them do not appear in the final expression of the determinant. Therefore, each term appearing in the final expression of the determinant of $A(p)+B(p)F(\tilde{p})C(p)$ is a signed product of distinct parameters and corresponds to a multi-colored cycle subgraph of $\mathbb{G}_F$.

Two multi-colored cycle subgraphs of $\mathbb{G}_F$ are similar if their corresponding terms are the product of the same set of parameters, but possibly with opposite signs. The sign of a term in the determinant remains unchanged if each factor of the form $p_r \tilde{p}_{\tilde{s}} p_t$ is replaced by a new parameter, and the number of cycles in a multi-colored cycle subgraph remains unchanged if a directed path from state vertex $x_j$ to state vertex $x_i$ by going through input and output vertices is replaced by an arc from $x_j$ to $x_i$. By Proposition 4 in \cite{liu2019graphical}, the sign of a term in the determinant of $A(p)+B(p)F(\tilde{p})C(p)$ is positive if $n-c$ is even, and is negative if $n-c$ is odd, where $c$ is the number of cycle graphs in the corresponding multi-colored cycle subgraph of $\mathbb{G}_F$. So a similarity class of multi-colored cycle subgraphs is unbalanced if among the corresponding terms of the graphs in the class, the number of positive terms and the number of negative terms are unequal, i.e., the corresponding terms do not cancel each other out. Therefore, $\mathbb{G}_F$ has an unbalanced similarity class of multi-colored cycle subgraphs if and only if the determinant is nonzero, which means the matrix $A(p)+B(p)F(\tilde{p})C(p)$ has full generic rank. \hfill $\qed$

\noindent \textbf{Proof of Theorem \ref{thm:binary}:} By Theorem \ref{thm:lnr}, (\lowercase\expandafter{\romannumeral1}) $\implies$ (\lowercase\expandafter{\romannumeral2}). The inverse will be proved by contradiction. Now assume (\lowercase\expandafter{\romannumeral2}) is true but (\lowercase\expandafter{\romannumeral1}) is false. That is, assume $\grank (A(p)+B(p)F(\tilde{p})C(p)) = n$, there are no subset $\mathcal{S} \subset \mathbf{k}$ and no permutation matrix $\Pi$ bringing the triple $(C_{\mathbf{k} - \mathcal{S}}(p), A(p), B_{\mathcal{S}}(p))$ into the form (\ref{eqn:ABC-partition}), and the system $\{A(p), B_i(p), C_i(p); k\}$ is structurally incomplete. By Theorem \ref{thm:lnr}, structural incompleteness implies that $\exists \mathcal{S} \subset \mathbf{k}$ such that $\forall p \in {\rm I\!R}^q$, $\left\langle A \, | \, \mathscr{B}_{\mathcal{S}} \right\rangle$ is a proper subspace of $[C_{\mathbf{k} - \mathcal{S}} \, | \, A]$. It is well known that $\left\langle A \, | \, \mathscr{B}_{\mathcal{S}} \right\rangle \subset [C_{\mathbf{k} - \mathcal{S}} \, | \, A]$ is equivalent to $C_{\mathbf{k} - \mathcal{S}}A^jB_{\mathcal{S}}=0$ for all $j \geq 0$. The binary assumption means that the coefficients of the parameters in the entries of $A(p)$, $B_{\mathcal{S}}(p)$, and $C_{\mathbf{k} - \mathcal{S}}(p)$ are all 1's, so the nonzero entries in these matrices cannot cancel each other out in matrix multiplication as they all have positive signs. Thus the fact that $C_{\mathbf{k} - \mathcal{S}}A^jB_{\mathcal{S}}=0$ for all $j \geq 0$ implies that there exists a permutation matrix $\Pi_1$ such that
\begin{align*}
& \Pi_1 A(p) \Pi_1^{-1} = 
\begin{bmatrix}
A_{11} & \mathbf{0} \\
A_{21} & A_{22}
\end{bmatrix}, \quad
\Pi_1 B_{\mathcal{S}}(p) = 
\begin{bmatrix}
\mathbf{0} \\
B_2
\end{bmatrix} \\
& C_{\mathbf{k} - \mathcal{S}}(p) \Pi_1^{-1} = 
\begin{bmatrix}
C_1 & \mathbf{0}
\end{bmatrix}
\end{align*} 
The assumption that there is no permutation matrix $\Pi$ bringing the triple $(C_{\mathbf{k} - \mathcal{S}}(p), A(p), B_{\mathcal{S}}(p))$ into the form (\ref{eqn:ABC-partition}) implies that both pairs $(A_{22}, B_2)$ and $(A_{11}', C_1')$ are irreducible. As $\grank (A(p)+B(p)F(\tilde{p})C(p)) = n$, by Lemma \ref{lem:grk-acc}, $\exists p \in {\rm I\!R}^q$ for which $\left\langle A \, | \, \mathscr{B}_{\mathcal{S}} \right\rangle = [C_{\mathbf{k} - \mathcal{S}} \, | \, A]$. This is a contradiction to the fact that $\forall p \in {\rm I\!R}^q$, $\left\langle A \, | \, \mathscr{B}_{\mathcal{S}} \right\rangle$ is a proper subspace of $[C_{\mathbf{k} - \mathcal{S}} \, | \, A]$. Thus, (\lowercase\expandafter{\romannumeral2}) $\implies$ (\lowercase\expandafter{\romannumeral1}).

By Theorem 4 in \cite{pichai1984graph} and its proof, there exist no subset $\mathcal{S} \subset \mathbf{k}$ and no permutation matrix $\Pi$ such that (\ref{eqn:ABC-partition}) holds if and only if each state vertex is in some strongly connected component (SCC) of the feedback graph $\mathbb{G}_F$, which contains an arc from $\mathcal{E}_F$. That is, each SCC of $\mathbb{G}_F$ which has at least one state vertex also contains an input vertex and an output vertex. If a SCC of $\mathbb{G}_F$ consists solely of input vertices and/or output vertices, then the SCC has exactly one vertex, because there are no arcs within input vertices, within output vertices, or from an input vertex to an output vertex in $\mathbb{G}_F$. So each SCC of $\mathbb{G}_F$ consists of either an input or output vertex or at least one input vertex, one state vertex, and one output vertex. By Lemma \ref{lem:grk-graph}, $\grank (A(p)+B(p)F(\tilde{p})C(p)) = n$ if and only if the feedback graph $\mathbb{G}_F$ has an unbalanced similarity class of multi-colored cycle subgraphs. Therefore, (\lowercase\expandafter{\romannumeral2}) $\iff$ (\lowercase\expandafter{\romannumeral3}). \hfill $\qed$

\section{Concluding Remarks}

This paper establishes algebraic conditions for the structural completeness of polynomially parameterized and linearly parameterized multi-channel systems respectively. This paper also gives an equivalent graphical condition for the class of linearly parameterized multi-channel systems satisfying the binary assumption. Some future research problems are: (1) to determine the complexity class of checking the graphical condition in Theorem \ref{thm:binary}; (2) to study some design problems of a linearly parameterized multi-channel system with the requirement of structural completeness; for example, given the graph $\mathbb{G}$ of a linearly parameterized system $(C(p), A(p), B(p))$, a design problem is to find the minimum number of feedback arcs (from the output vertices to the input vertices) that can be added to $\mathbb{G}$ such that the resulting graph is the feedback graph $\mathbb{G}_F$ of a structurally complete multi-channel system; (3) to extend the graph-theoretic condition in Theorem \ref{thm:binary} to all linearly parameterized multi-channel systems using weighted directed multigraphs.

\section*{Acknowledgements}

We would like to thank Yifeng Liu for a rigorous proof of Lemma \ref{lem:partition} and Brian D. O. Anderson for his valuable feedback.


\bibliographystyle{IEEEtran}
\bibliography{fjbib-fix-spec}


%
%
%
%
%
%
%
%
%
%
%
%


\end{document}